\documentclass{aastex63} 

\accepted{August 30, 2021}

\submitjournal{AJ}

\usepackage{apjfonts}

\def\lae{\mathrel{\raise .4ex\hbox{\rlap{$<$}\lower 1.2ex\hbox{$\sim$}}}}
\def\gae{\mathrel{\raise .4ex\hbox{\rlap{$>$}\lower 1.2ex\hbox{$\sim$}}}}

\newcommand{\axaf}{\mbox{\em Chandra\/}}

\newcommand{\suzaku}{\mbox{\em Suzaku\/}}
\newcommand{\nustar}{\mbox{\em NuSTAR\/}}

\newcommand{\xmm}{\mbox{\em XMM-Newton\/}}
\newcommand{\NeX}{Ne {\sc x}}
\newcommand{\OVIII}{O {\sc viii}}
\newcommand{\FeXVII}{Fe {\sc xvii}}

\newcommand{\bmean}[1]{\beta_#1}
\newcommand{\hatbmean}[1]{\hat\beta_#1}

\newcommand{\bvar}[1]{\Lambda_#1}
\newcommand{\hatbvar}[1]{\hat{\Lambda}_#1}
\newcommand{\bcor}{\rho_{uv}}
\newcommand{\hatbcor}{\hat\rho_{uv}}

\usepackage{amsmath}
\usepackage{comment}
\usepackage{appendix}
\shorttitle{Demonstrating X-ray Calibration Concordance}
\shortauthors{Marshall et al.}

\begin{document}

\title{Concordance: In-flight Calibration of X-ray Telescopes without Absolute References}

\correspondingauthor{Herman L.\ Marshall}
\email{hermanm@space.mit.edu}

\author{Herman L.\ Marshall}
\affiliation{Kavli Institute for Astrophysics and Space Research,
 Massachusetts Institute of Technology, 77 Massachusetts Ave.,
 Cambridge, MA 02139, USA}
 
\author{Yang Chen}
\affiliation{U. Michigan, Ann Arbor, MI 48109, USA}

\author{Jeremy J. Drake}
\affiliation{Harvard-Smithsonian Center for Astrophysics,
 60 Garden St., Cambridge, MA 02138, USA}
 
\author{Matteo Guainazzi}
\affiliation{ESTEC, Keplerlaan 1 2201AZ Noordwijk,
 The Netherlands}

\author{Vinay L.\ Kashyap}
\affiliation{Harvard-Smithsonian Center for Astrophysics, 60 Garden St., Cambridge, MA 02138, USA}
 
\author{Xiao-Li Meng}
\affiliation{Harvard University, Cambridge, MA, 02138, USA}

\author{Paul P. Plucinsky}
\affiliation{Harvard-Smithsonian Center for Astrophysics,
 60 Garden St., Cambridge, MA 02138, USA}

\author{Peter Ratzlaff}
\affiliation{Harvard-Smithsonian Center for Astrophysics,
 60 Garden St., Cambridge, MA 02138, USA}
 
\author{David A.\ van Dyk}
\affiliation{Imperial College, London, UK SW7 2AZ}
 
\author{Xufei Wang}
\affiliation{Harvard University, Cambridge, MA, 02138, USA}

\begin{abstract}

We describe a process for cross-calibrating the effective areas of X-ray telescopes that observe common targets.
The targets are not assumed to be ``standard candles'' in the classic sense,
in that
we assume that the source fluxes have well-defined, but {\sl a priori} unknown values. 
Using a technique developed by \citet{concordancejasa} that involves a statistical
method called {\em shrinkage estimation}, we determine effective area correction factors
for each instrument that brings estimated fluxes into the best agreement,
consistent with prior knowledge of their effective areas.
We expand the technique to allow unique priors on systematic uncertainties in
effective areas for each X-ray astronomy instrument and to allow correlations
between effective areas in different energy bands.
We demonstrate the method with several data sets from various X-ray telescopes.

\end{abstract}


\section{Introduction}

We address a perennial issue in instrument performance, when estimated fluxes using two or more instruments
disagree.
While many instruments can safely rely on calibration traceable to established standards,
the performance of a space-based telescope usually cannot be recalibrated, as the instruments
are not returned to the lab.
Furthermore, space-based instruments may be affected by the physical rigors of launch into space and
instrument performance can change with time due to gas leakage, filter deterioration, component failure,
contamination buildup and other reasons.
Without absolute standards that may be observed while in space, astronomers
generally resort to the use of secondary, astronomical standards that can be observed
while operating.
Nonvariable sources are generally chosen as secondary standards so that they can be
reused by the telescope team and observed by others.
For example, in the 2-8 keV energy range, the Crab Nebula was frequently
used as a standard, especially the observation by \citet{1974AJ.....79..995T}; for a recent use of
the Crab Nebula for cross-calibrating X-ray telescopes, see \citet{2005SPIE.5898...22K} for
a comparison between many missions and \citet{2017ApJ...841...56M} for a case of
assessing the \nustar\ telescope effective model.
However, most astronomical sources vary, even the Crab Nebula, at levels
detectable by current instruments, necessitating another approach:
joint in-flight observations for cross-calibration of instruments.

Our work is set upon the foundation established by the International
Astronomical Consortium for High Energy Calibration (IACHEC).
The IACHEC was formed primarily to assist X-ray telescope teams
who cross-calibrate instruments and to understand
the sources used for this purpose.
See recent IACHEC reports for summaries of
recent activity \citep{iachec2019,iachec2020}.
All of the IACHEC working groups address
issues of cross calibration of X-ray telescopes and often find
discrepancies between results for the same source.
Examples of IACHEC
work include observations of the supernova remnants (SNRs) G21.5 \citep{2011A&A...525A..25T} and
1E 0102$-$7219 \citep{2017A&A...597A..35P},
spectra of galaxy clusters \citep{2010A&A...523A..22N,2013A&A...552A..47K,2015A&A...575A..30S},
spectra of white dwarfs and isolated neutron stars \citep{2006A&A...458..541B},
and simultaneous observations of active galaxies such as 3C 273
and PKS 2155$-$304 \citep{2011PASJ...63S.657I,2017AJ....153....2M}.
The studies had a common
problem: assessing how much any particular instrument's effective
area should be adjusted so that the measurements might agree.
Ordinary weighting of measurements based on photon counting statistics would
give the observations with large effective areas and exposure times
the greatest influence on the result but without consideration of possible
systematic errors.
{The overarching goal of IACHEC is thus to bring the competing adjustments to instrumental effective areas into concordance, while simultaneously including prior knowledge of possible systematic errors.  

Here, we further develop the ``shrinkage'' method pioneered by \citet[][]{concordancejasa}, hereafter referred to as Paper I, to compute objective corrections to effective areas of several high-energy instruments.
See Section~\ref{sec:background} for the model setup and notation.
The method is extended to account for systematic uncertainties specific to each instrument and incorporate systematic correlations across passbands.  We present the method here in some generality, without the details of the mathematical machinery presented in Paper~I 
(Section~\ref{sec:method}), along with extensions added (Section~\ref{subsec:extension}), and apply it to a variety of datasets (Section~\ref{sec:data}). We present the results in Section~\ref{sec:results} along with simulation studies that aim to validate the method, and discuss the next steps in Section~\ref{sec:summary}.
We note that while we examine the case of X-ray telescopes in this paper, the method is extendable to most types of telescopes.

\section{Method}

\subsection{Background}\label{sec:background}

\noindent
We start with an idealized calibration data set where $M$ objects are observed by each of $N$ instruments,
obtaining photon counts, $c_{ij}$, where $j=1,\ldots, M$ indexes the
sources and $i=1,\ldots, N$ indexes the instruments.\footnote{We consider each unique combination
of telescope and detector with any filters or gratings to be an ``instrument''.} Each observation is described by a set of observational parameters (e.g., exposure time) that we encapsulate in a matrix {$\mathbf{T}=\{T_{ij}\}$}.

Denoting the true effective area of instrument $i$ by $A_i$ and the true flux of source $j$ by $F_j$,
the expected counts $C_{ij}$ for each object/detector combination is given by

\begin{equation}
\label{eqn:multiplicative}
C_{ij} = T_{ij} A_i F_j, \quad 1\leq i\leq N,\quad 1\leq j\leq M,
\end{equation}

\noindent
where $T_{ij}$ has units of seconds $\times$ counts per photon, $F_{j}$ has units of photons
per unit area per second, and $A_i$ has units of area. We assume for now that the true exposure factors have negligible error and equal the observed values, i.e., $T_{ij} = t_{ij}$. We follow the
notation of Paper I by using lower case to indicate measured quantities and upper case
to indicate the ``true'' values to be estimated. Note that in fact, the multiplicative constant $T_{ij}$ contains not only the exposure time
but also other factors that can be calculated precisely for any given observation.

We aim to estimate $F_j$ using the calibration data, i.e., the observed counts $c_{ij}$, and an external (prior) estimated effective area, $a_i$ for each instrument. A naive procedure simply substitutes each quantity in Eq.~\eqref{eqn:multiplicative} with its measured counterpart, to obtain the ``estimating equation'',
\begin{equation}
c_{ij} = t_{ij} a_i f_{ij}
\label{eq:data}
\end{equation}
and solving Eq.~\eqref{eq:data} for $f_{ij}$ for each instrument-source combination, thus yielding $N$ different estimates of each flux. 
The resulting estimator $f_{ij}= c_{ij}/(a_i t_{ij})$ is a \textit{ratio estimator}, which is known in statistical literature to be both seriously biased and highly variable. More precisely, the variability in the denominator can cause large uncertainty (considering dividing by a value close to zero). Furthermore, the average,
$\frac{1}{N} \sum_{i=1}^N f_{ij}$, is a biased estimator of $F_j$.
We return to the issue associated with ratio estimators in \S\ref{sec:posteriors}.

We can do much better by analyzing all data together using more principled and sophisticated statistical methods, such as the one given in Paper I. We can then achieve our goal to obtain {\em better} estimates of the instrumental effective areas, $A_i$, that bring our flux estimates, $f_{ij}$, closest to the $F_j$ in some strict statistical sense. Fig.~\ref{fig:schematic} gives a schematic representation of our goal.

\begin{figure}[t!]
\includegraphics[width=10cm]{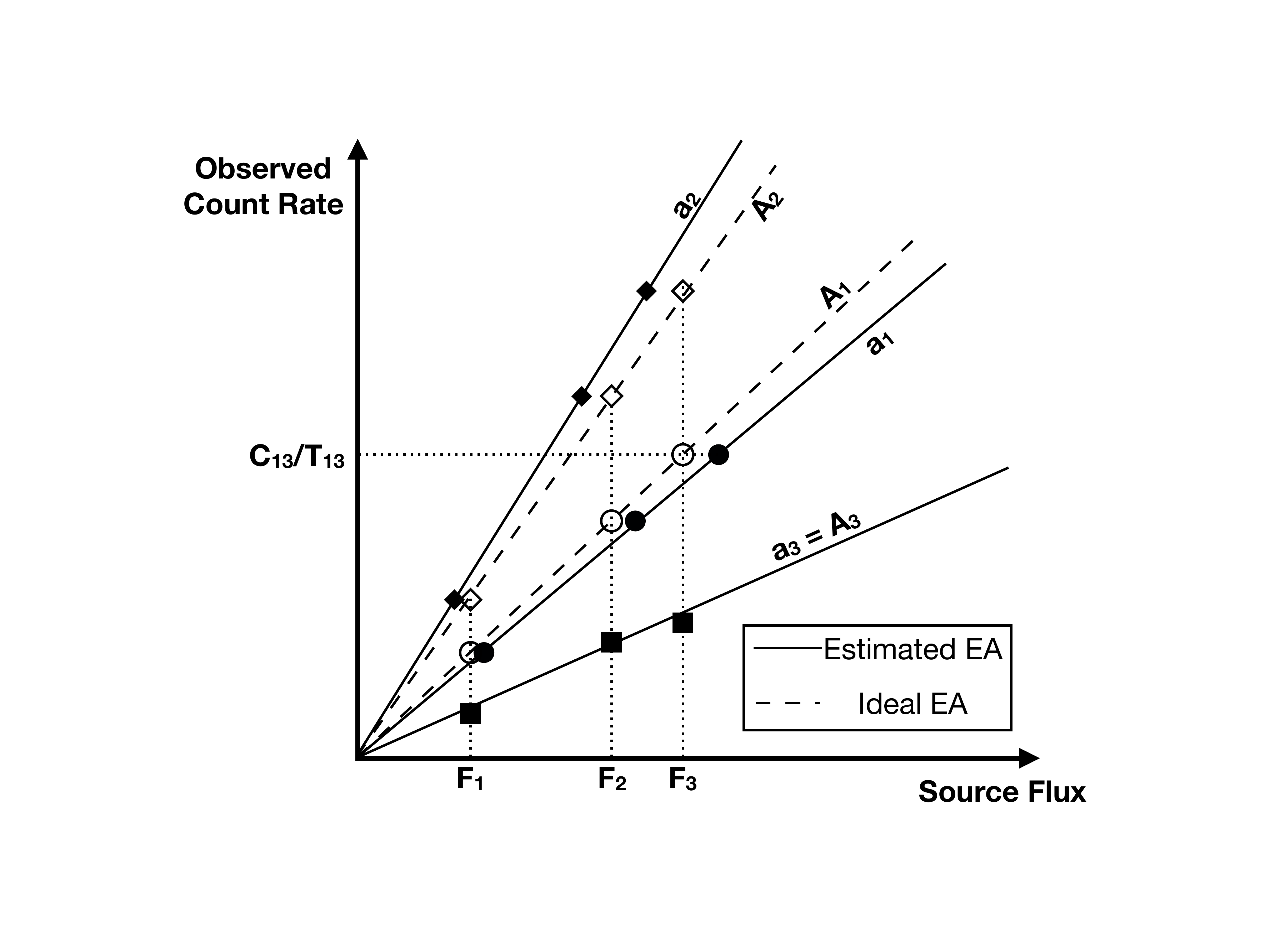}
\caption{
 The goal of Concordance illustrated schematically.  This schematic supposed $3$ sources and $3$ instruments, and plots expected count rates,
 $C_{ij}/T_{ij}$, on the vertical axis and plots true and estimated fluxes on the horizontal axis. The three instruments are represented by different symbols, with solid symbols representing the naive estimates of $F_j$, i.e., $f_{ij} = c_{ij}/(T_{ij} a_{i})$, and open symbols representing the same, but with $a_i$ replaced by the true effective area, $A_i$.   The estimates are systematically biased because the $a_i$ are inaccurate estimates of $A_i$; compare the solid and dashed lines. (In the plot EA is used as an abbreviation for effective area.) Our aim is to estimate the $A_i$ values that yield best agreement between instruments for each source.
\label{fig:schematic} }
\end{figure}

In practice, 
estimating a flux from an observation is not as simple an operation as
merely counting events.
A strictly correct approach takes into consideration the response of the instrument,
especially when attempting to measure the flux in a specific bandpass.
It is often necessary to
use a forward-folding method, such as implemented in {\tt xspec},
that takes a count spectrum as input and effective area and
response functions as externally (and accurately) provided.
Here, we approximate this process using Eq.~\eqref{eqn:multiplicative}
because the flux (as defined here) is often a robustly computed quantity for
any given observation.  In fact, we do not actually compute $c_{ij}$
but a proxy for $c_{ij}$ as described below.

We now carefully examine the flux measurement process to
justify using the simplistic model represented by Eq.~\eqref{eqn:multiplicative}.
Source $j$ is assumed to have a photon spectrum $n_E(\Theta_j)$ in units
of photons per unit area per unit energy at energy $E$.  Each $\Theta_j$ represents a vector of
spectral parameters for a source, which we assume to include an overall normalization
$n_j$ with the same units as $n_E$, so $q(E; \theta_j) \equiv n_E(\Theta_j)/n_j $ defines the spectral
shape as a function of the remaining spectral parameters, $\theta_j$.
In the case of a spectrometer measuring an emission line, the bandpass may be so narrow that
$q$ is well approximated by a delta function, in which case $F_j = n_j$.
Otherwise, the flux in a band from $E_1$ to $E_2$ is
\begin{equation}
F_j = \int_{E_1}^{E_2} n_E(\Theta_j) dE = n_j \int_{E_1}^{E_2} q(E; \theta_j) dE
\end{equation}

\noindent
giving
\begin{equation}
\label{eq:nj}
n_j = \frac{F_j}{\int_{E_1}^{E_2} q(E; \theta_j) dE} \equiv \frac{F_j}{\tilde{q}_j \Delta E }
\end{equation}

\noindent
where $\Delta E = E_2 - E_1$.

For most X-ray telescopes, the instrumental effective area can be separated
into two parts.  There is the geometric area of the optics, i.e., $A_i^g$,  with
losses due to mechanical obscuration, reflections, and transmissions
of the optics (including filters) given by $r_i(E)$.  The other part is due to the
quantum efficiency of the detector, $Q(E)$, which gives probability of a detecting
a photon.  We characterize the (true) effective area of instrument $i$ as
a function of energy in the bandpass of interest
by {$\tilde{A}_i(E) = A_i^g r_i(E) Q_i(E) \equiv A_i  \alpha_i(E)$, where
$\alpha_i(E)$ now gives the
shape of the effective area in the band and is defined such that its integral
over the band is unity, i.e.,
\begin{equation}
    \alpha_i(E)=\frac{r_i(E) Q_i(E)}{\int_{E_1}^{E_2} r_i(E) Q_i(E) dE}. 
\end{equation}
Consequently,
\begin{equation}
    A_i = A_i^g \cdot \int_{E_1}^{E_2} r_i(E) Q_i(E) dE
\end{equation}
is the scale of the effective area and does not depend on $E$}. 
We take this approach because the model of the effective
area through the band is generally better 
known than its absolute value.

The detector provides counts in $K$ channels that are
related to the true photon energy, $E$, via a response function $\Phi_k(E)$, where

\begin{equation}
\sum_{k=1}^K \Phi_k(E) = 1.
\end{equation}

\noindent
A given observation has an accurately determined exposure time $t_{ij}$
that sets the expected count in channel $k$:

\begin{eqnarray}
      \label{eq:countijk}
C_{ijk} & = & t_{ij} \int \tilde{A_i}(E) n_E(\Theta_j) \Phi_k(E) dE \\
           & = & t_{ij} A_i n_j \int \alpha_i(E) q(E; \theta_j) \Phi_k(E) dE \\
           & = & t_{ij} A_i F_j \frac{\int \alpha_i(E) q(E; \theta_j)  \Phi_k(E) dE}{\tilde{q}_j \Delta E} \label{eq:intermediate},\\
           & \equiv & \tilde{T}_{ijk} A_i F_j,
\end{eqnarray}

\noindent
where Eq.~\eqref{eq:intermediate} follows from Eq.~\eqref{eq:nj}. It is important to point out that Eq.~\eqref{eq:countijk} is well-known in the astronomy literature but Eq.~\eqref{eq:intermediate} is an innovative way of simplifying the expressions, which is essential for tackling the current problem. Finally, the counts that are associated
with the energy band $(E_1, E_2)$ are mostly in the a range of channels
$(k_1, k_2)$, which is chosen so that we have a reliable estimate of flux in the band of interest,
giving the expected count relevant to a particular flux measurement  
\begin{equation}
C_{ij} = \sum_{k=k_1}^{k_2} C_{ijk} = A_i F_j \sum_{k=k_1}^{k_2} \tilde{T}_{ijk} \equiv T_{ij} A_i F_j,  \label{eq:meanmodel}
\end{equation}

\noindent
which defines $T_{ij}$ in terms of $t_{ij}$, a normalized sum over the response function,
and the two shape functions. This is the formal 
basis for Eq.~\eqref{eqn:multiplicative}.

In actual data analysis, the observed counts, $c_{ijk}$, not the expected counts, $C_{ijk}$, are input into the iterative routine that estimates the fluxes, $f_{ij}$, and, again, the estimated effective area, $a_i$, must be used because the true effective area is unknown.  Thus, in analogy to Eq.~\eqref{eq:data}, we replace $C_{ij}$, $A_i$, and $F_j$ in Eq.~\eqref{eq:meanmodel} with their observed counterparts to obtain the estimating equation

\begin{equation}
\label{eq:observedcij}
    c_{ij} = \sum_{k=k_1}^{k_2} c_{ijk} = 
     a_i f_{ij} T_{ij} = a_i f_{ij} \sum_{k=k_1}^{k_2} \tilde{T}_{ijk}.
\end{equation}

\noindent
Finally, naive, instrument-specific flux estimates are obtained by solving
Eq.~\eqref{eq:observedcij},

\begin{equation}
    f_{ij} = \frac{c_{ij}}{T_{ij} a_i},
\end{equation}
which is essentially a full derivation of the ratio estimator.

It is important to our analysis that the $T_{ij}$ values be independent of $A_i$ and $F_j$.
There are two circumstances that may cause a problem
in this regard: 1) when there is some nonlinearity
of the detector response, such as pileup, where $Q_i(E)$ depends on the
source flux, and 2) when $\alpha_i(E)$ is highly
nonlocal and the bandpass of integration in Eq.~\eqref{eq:countijk} is large, involving portions
of the spectrum where $q(E;\theta_j)$ or $\alpha_i(E)$ are poorly determined.
By choosing instrument data in the linear regime, avoiding pileup, and restricting
the bandpasses of interest, we mitigate these potential issues.

The goal of our statistical modeling is to determine the best estimates of $A_i$ and $F_j$
consistent with the data, $c_{ij}$ with uncertainties $\sigma_{ij}$.
If all the effective areas were precisely correct, i.e. if $A_i = a_i$, then 
we could estimate the $F_j$ via a relatively trivial regression model. 
However, we  know that there are systematic errors in our estimated effective areas because observations show that the $f_{ij}$ do not cluster about any given value within their individual uncertainties.
We addressed this problem in Paper I by introducing estimates of the systematic errors on the estimated effective areas and applying a statistical method called {\em shrinkage estimation}.

\subsection{Statistical Model and Shrinkage Estimates.}

\label{sec:method}
Paper I proposes a linear regression model for the log count rates, denoted 
$y_{ij} \equiv \log (c_{ij}/T_{ij})$, such that
\begin{equation}
    y_{ij} = B_i + G_j - \frac{1}{2 \sigma_i^2} + e_{ij},
    \label{eq:jasa}
\end{equation}
where the $B_i \equiv \log A_i$, and the $G_j \equiv \log F_j$ are 
the quantities to be estimated, and the $e_{ij}$ are 
noise terms that are assumed normally distributed with mean $0$ and (known) variance $\sigma_i^2$. The 
{$-\frac{1}{2\sigma_i^2}$}
term in Eq.~\eqref{eq:jasa} is a half-variance correction that is included to maintain the multiplicative mean modeling in Eq.~\eqref{eq:meanmodel}.
This correction ensures that $\hbox{E}(c_{ij}) = C_{ij} = T_{ij} A_i F_j$ because if $\log x \sim \hbox{N}(\mu, v)$, then $\hbox{E}(x) = e^{0.5v+\mu}$.

Paper I adopts a Bayesian hierarchical modeling approach to estimate the unknown quantities $B_i$, $G_j$, and $\sigma_i^2$. This involves setting independent Gaussian prior distributions for the $B_i$, with prior mean $b_i=\log a_i$ and prior variance $\tau_i^2$ and setting independent flat priors for the $G_i$; and setting independent conjugate priors for the $\sigma_i^2$.
Paper I goes on to show how a Hamiltonian Monte Carlo algorithm can be used to obtain a sample from the joint posterior distribution of all unknown quantities in Model~\eqref{eq:jasa} and cross-checks this computation with a blocked Gibbs sampler. The resulting Monte Carlo sample can be transformed back to the effective areas and source fluxes on their original scale to obtain their posterior distributions, estimates, and error bars. 

A Bayesian perspective allows us to combine multiple sources of information -- in this case the information from the calibration data, the $y_{ij}$, and from the prior estimates of the effective areas, the $a_i$. 
By replacing the estimated effective areas with prior distributions that reflect their uncertainty, we are able to update the estimated effective areas and their error bars in light of the calibration data. This approach provides improved estimates (e.g., in terms of mean squared error) of the effective areas and the fluxes
simultaneously.  As a result, we obtain estimates of the sources' true fluxes
that combine the instrument-specific estimates in a statistically principled manner.  

The updated estimates of the effective areas are weighted averages of their priors and the best values based on the current calibration data. In the context of
{Model~\eqref{eq:jasa}},
we work on the log scale.  The estimates of ${B}_i$ and ${G}_j$ are given by

\begin{equation}
    \widehat{B}_i 
    = W_i (\bar{y}_{i\cdot}' - \bar{G}_i)+(1-W_i) b_i \quad \text{ and } \quad  \widehat{G}_j 
    = \bar{y}_{\cdot j}' - \bar{B}_j,
    \label{eq:shrink}
\end{equation}
where

\begin{equation}
\bar y_{i\cdot}'=\frac{\sum_{j =1}^M y'_{ij}\sigma^{-2}_{i}}{\sum_{j=1}^M\sigma^{-2}_{i}}, 
\quad
\bar y_{\cdot j}'=\frac{\sum_{i=1}^N y'_{ij}\sigma^{-2}_{i}}{\sum_{i=1}^N\sigma^{-2}_{i}}, 
\quad
\bar G_{i}=\frac{\sum_{j=1}^M\widehat G_{j}
\sigma^{-2}_{i}}{\sum_{j=1}^M\sigma^{-2}_{i}}, 
\quad
\bar B_{j}=\frac{\sum_{i=1}^N\widehat B_{i}
\sigma^{-2}_{i}}{\sum_{i=1}^N\sigma^{-2}_{i}}, 
\end{equation}

\noindent
$y'_{ij}={y}_{ij}+0.5\sigma^2_i$, and the weights are given by $W_i= M\sigma^{-2}_{i}/(\tau^{-2}_i+M\sigma^{-2}_{i})$.
If the prior estimate of the effective area of a particular instrument is very precise relative to its calibration data, i.e., $\tau_i^2 \ll \sigma_i^2/M$,
then
$W_i\approx 0$, and the updated estimate of that instrument's
effective area is dominated by the prior estimate, resulting in $\widehat B_i\approx b_i$.
In contrast, if the calibration data are much more precise, then the weights are near
unity and the updated estimate of the effective area is
dominated by the calibration data, giving $\widehat B_i\approx \bar{y}_{i\cdot}' - \bar{G}_i$.

The estimates $\widehat B_i$
in Eq.~\eqref{eq:shrink} are often called ``shrinkage estimates'' due to their historical use for  ``shrinking" several estimates together toward a common prior mean~\citep{efron1972empirical,efron1973stein} when, for example, the $b_i$ are all the same.
Because the prior means, $b_i$, are different for different instruments,
the $\widehat B_i$ are simply a sensible combination
of prior knowledge captured by $b_i$ and data represented by $\bar{y}_{i\cdot}'$,  weighted by their respective precisions, which are the reciprocals of their variances (assuming the $\sigma_i$ are known). This combination
allows our model to weigh the prior estimate of the effective area
for a given instrument
against deviations between the observed fluxes of the same sources from different instruments and ultimately to obtain the joint estimates of the true fluxes and effective areas that are most consistent with the calibration data and the priors on the effective areas. 

Paper~I further describes how to handle the case where all sources are not observed by all instruments and presents a robust version of Model~\eqref{eq:jasa} that allows for outliers among the measured fluxes (or source counts) by replacing the $\log$ Normal error model with a $\log t$ error model.

\subsection{Extensions of the Model}
\label{subsec:extension}

Paper I proposes 
modeling calibration data using
{Model}~\eqref{eq:jasa}
and its extensions, derived computational methods for fitting these models, and validated their statistical properties.
However, application of {Model}~\eqref{eq:jasa} to real data
requires relaxing some of its basic assumptions.
Here, we illustrate how this is accomplished via two extensions to the method.

\subsubsection{Heterogeneous Uncertainties in Effective Area Priors}

\label{sec:hetero}

IACHEC scientists recognize that the quality of ground-based calibrations
varies significantly from instrument to instrument, resulting in perceived differences
in the reliabilities of the estimated effective areas.
The formalism laid out in Paper I allows for instrument specific prior distributions for
for the $B_i$, as explained in Section~\ref{sec:method}, given by
Gaussian distributions
with instrument-specific 
variances $\tau_i^2$.
In the numerical examples in Paper~I, however, the $\tau_i^2$ were set to $\tau^2$ for each $i$,
essentially
assuming that all modelled calibrations are equally uncertain in percentage terms.
Here we allow for {\em heterogeneous} $\tau^2_i$ values, as covered by the theory given in Paper I.
At IACHEC meetings in 2017, 2018, and 2019 \citep{iachec2019,iachec2020},
we asked instrument calibration scientists
to specify values of $\tau_i$ for their instruments in each of a specific set of bandpasses.
These values are given in Tables~\ref{tab:tau1} and \ref{tab:tau2}; instruments with
significant effective area below 1 keV appear in Table~\ref{tab:tau1} and other instruments appear in Table~\ref{tab:tau2}.

Of course, in practice it is difficult even for experts to quantify the $\tau_i$ precisely. Thus, it is important that we examine the sensitivity of our results to the specified values.
Often, however, there is a body of experience and expert knowledge on the reliability of ground-based standards that allows rough estimation of systematic errors.

\begin{deluxetable}{lcccccccccc}
\tablecolumns{10}
\tablewidth{0pc}
\tablecaption{Effective Area Uncertainty Priors
    ($\tau_i$)\tablenotemark{a} \label{tab:tau1} }
\tablehead{ \colhead{} & \multicolumn{9}{c}{Energy Bands (keV)} \\
\colhead{Instrument} &
	\colhead{0.15-0.33}	&	\colhead{0.33-0.54}	&	\colhead{0.54-0.8}	&	\colhead{0.8-1.2} &
	\colhead{1.2-1.8}	&	\colhead{1.8-2.2}	&
    \colhead{2.2-3.5}	&	\colhead{3.5-5.5}	&	\colhead{5.5-10} }
\startdata
Astrosat SXT	&	\nodata	&	15	&	15	&	10	&	10	&	10	&	10	&	10	&	10	\\	
Chandra ACIS	&	3	&	3	&	3	&	3	&	2.6	&	3.3	&	3.3	&	4.9	&	5	\\
Chandra HETGS	&	\nodata	&	\nodata	&	10	&	5	&	4	&	4	&	4	&	5	&	7	\\	
Chandra LETGS	&	5	&	7	&	7	&	7	&	7	&	7	&	7	&	10	&	10	\\	
ROSAT PSPC		&	10	&	10	&	10	&	10	&	10	&	10	&	\nodata	&	\nodata	& \nodata		\\	
Suzaku XIS1		&	\nodata	&	20	&	15	&	10	&	10	&	15	&	5	&	5	&	5	\\	
Suzaku XIS0,2,3		&	\nodata	&	\nodata	&	15	&	10	&	10	&	15	&	5	&	5	&	5		\\	
Swift PC/WT		&	\nodata	&	15	&	10	&	7.5	&	7.5	&	10	&	5	&	5	&	5	\\	
XMM MOS1,2		&	20	&	10	&	6	&	6	&	6	&	6	&	6	&	6	&	10	\\	
XMM pn			&	2	&	2	&	2	&	2	&	2	&	2	&	2	&	2	&	3	\\	
XMM RGS	        &	\nodata	&	8	&	5	&	5	&	5	&	\nodata	&	\nodata	& \nodata & \nodata \\
\enddata
\tablenotetext{a}{The $\tau_i$ values are given as percentages.  The ellipses indicate
bandpasses where the instrument has an negligible effective area.}
\end{deluxetable}

\begin{deluxetable}{lcccccccc}
\tablecolumns{8}
\tablewidth{0pc}
\tablecaption{Effective Area Uncertainty Priors
    ($\tau_i$)\tablenotemark{a} \label{tab:tau2} }
\tablehead{ \colhead{} & \multicolumn{7}{c}{Energy Bands (keV)} \\
\colhead{Instrument} &
    \colhead{2.2-3.5}	&	\colhead{3.5-5.5}	&	\colhead{5.5-10}	&
	\colhead{15-25}	&	\colhead{25-50}	&	\colhead{50-100}	&	\colhead{100-300} }
\startdata
Astrosat CZTI	&	\nodata	&	\nodata	&	\nodata	&	\nodata	20	&	20	&	20	&	25 \\
Astrosat LAXPC	&	\nodata	&	15	&	15	&	15	&	15	&	20	&	\nodata	\\	
INTEGRAL IBIS	&	\nodata	&	\nodata	&	\nodata	&	\nodata	&	8	&	15	&	20 \\
INTEGRAL SPI	&	\nodata	&	&	\nodata	&	\nodata	&	5	&	5	&	5 \\
NuSTAR	&	\nodata	&	4	&	3	&	3	&	15	&	20	&	\nodata	\\	
RXTE PCA	&	5	&	10	&	3	&	3	&	10	&	50	&	\nodata	\\
RXTE HEXTE	&	\nodata	&	\nodata	&	\nodata	&	5	&	5	&	5	&	\nodata	\\
Suzaku HXD	&	\nodata	&	\nodata	&	\nodata	&	20	&	20	&	20	&	20 \\
Swift BAT	&	\nodata	&	\nodata	&	\nodata	&	15	&	4	&	4	&	12 \\
\enddata
\tablenotetext{a}{The $\tau_i$ values are given as percentages.}
\end{deluxetable}

\subsubsection{Prior Correlations among Effective Areas}

\label{sec:correlation}
The second extension allows correlations between the
effective areas in different energy bands for each instrument.
In Paper I, we treated
different energy bands as separate (independent) instruments, while in reality
their effective areas can be strongly correlated.
Continuing to work on the log relative scale given in Eq.~\eqref{eq:transform}, 
we denote the effective areas as a function of the energy band $\mathcal{E}=[E_1,E_2]$ by ${B}(\mathcal{E}, \vec{\xi})=\log \int_{E_1}^{E_2} A(E, \vec{\xi})~d E$,
where $\vec{\xi}$ parameterizes the effective area and 
includes quantities such as the geometric area, filter thicknesses, and chemical compositions that are initially estimated during ground calibration. {Uncertainties in $\vec{\xi}$ are quantified via the prior distribution, $p(\vec{\xi})$; examples generated and used for ACIS analyses can be found in 
\citet{2006SPIE.6270E..1ID,2008SPIE.7016E..0PK,2011ApJ...731..126L} and \citet{2014ApJ...794...97X}.
}
We suppress the subscript $i$ throughout this section for notational simplicity because we are restricting consideration to an arbitrary instrument.

Following the discussion in Section~\ref{sec:method}, we specify the prior distribution on (the logarithm of) the effective areas of $U$ energy bands, $\{{B}(\mathcal{E}_u,\vec{\xi})\}$ for ${u=1,\ldots, U}$, as a multivariate Gaussian distribution with 
expected values, $\bmean{u}$, 
and variances, $\bvar{u}$, for each energy band $u$, and with correlations, $\bcor$, between all pairs of distinct energy bands, $u\neq v$.
Among the $b_u$, the $\Lambda_u$, and the $\rho_{uv}$,
only the correlations, $\rho_{uv}$, (or, more precisely, the Monte Carlo estimates of the $\rho_{uv}$) are used in our data analyses. To distinguish the Monte Carlo estimates of the prior means and variances of the $B_i$ obtained here from those we actually use, we introduce new notation for these quantities  that differ from  those used in Paper~I and in Section~\ref{sec:method}.
The current methods for computing $\Lambda_u$ are still
somewhat experimental, so we rely instead on the $\tau_i$ values from
IACHEC scientists.

For a given instrument and energy band $u$, the expectation is
\begin{equation}
    \bmean{u} = \int {B}(\mathcal{E}_u,\vec{\xi}) ~p(\vec{\xi}) ~d \vec{\xi}.
    \label{eq:bmean}
\end{equation}

\noindent
In practice, calibration scientists set $a_i$ to be the prior estimate of $A_i$ (on the original scale) based on their best information and experience. Transforming to a logarithmic scale, we might set our prior estimate of $B_i$ to {$\log a_i$}, 
which is the choice of prior mean we suggest in Section~\ref{sec:method}. Eq.~\eqref{eq:bmean} can be used in the absence of such intuition or if we prefer to use a parameterized model for ${B}$. (This is particularly relevant for the correlations, since we have less intuition for them.)  In this case, a reasonable strategy is to proceed via Monte Carlo integration of Equation~\eqref{eq:bmean}. More precisely, we obtain a {\it calibration sample}, $\{{B}^{(k)}(\mathcal{E}_u), k=1,\ldots,K\}$, that quantifies prior uncertainty in $B(\mathcal{E}_u,\vec{\xi})$, for example by obtaining a sample of size $K$ from $p(\vec{\xi})$ and computing $B(\mathcal{E}_u,\vec{\xi}^{(k)})$ for each sample, $\vec{\xi}^{(k)}$, from $p(\vec{\xi})$.
The $B^{(k)}(\mathcal{E}_u)$ is expressed as $\log \left[ \int_{E_1}^{E_2} A^{(k)}(E)~dE \right]$ for each $u$ (given all Monte Carlo samples $A^{(k)}(E)$ for $E\in\mathcal{E}_u$).  The Monte Carlo version of Eq.~\eqref{eq:bmean} is then

\begin{equation}
    \hatbmean{u} = \frac{1}{K} \sum_{k=1}^K {B}^{(k)}(\mathcal{E}_u).
    \label{eq:bmean-mcmc}
\end{equation}

\noindent
The prior variance of ${B}(\mathcal{E}_u,\vec{\xi})$ and its Monte Carlo estimate are

\begin{eqnarray}
    \bvar{u}  =  \int [ B(\mathcal{E}_u,\vec{\xi})-\bmean{u}]^2 ~p(\vec{\xi}) ~d \vec{\xi} 
 \quad \hbox{ and  } \quad
    \hatbvar{u}  =  \frac{1}{K-1} \sum_{k=1}^K \left[{B}^{(k)}(\mathcal{E}_u) - \hatbmean{u} \right]^2,
\end{eqnarray}

\noindent
respectively. Finally, the prior correlation between 
${B}(\mathcal{E}_u,\vec{\xi})$ and ${B}(\mathcal{E}_v, \vec{\xi})$ is the covariance normalized by the respective standard deviations, 

\begin{equation}
\label{eq:correlation}
    \bcor = \frac{1}{\sqrt{\bvar{u} \bvar{v}}}
        \int \left[ B(\mathcal{E}_u,\vec{\xi})-\bmean{u}\right]
        \left[\tilde{ B}(\mathcal{E}_v,\vec{\xi})-\bmean{v}\right] ~p(\vec{\xi}) ~d \vec{\xi}
\end{equation}

\noindent
with  Monte Carlo estimate,
\begin{equation}
\label{eq:correlation-hat}
\hatbcor = 
\frac{1}
    {(K-1)\sqrt{\hatbvar{u} \hatbvar{v}}} \ \
    \sum_{k=1}^K
\left[{B}^{(k)}(\mathcal{E}_u)-\hatbmean{u}\right]
        \left[{B}^{(k)}(\mathcal{E}_v)-\hatbmean{v}\right].
\end{equation}

\subsection{Practical Implementation}

This section discusses practical implementation of our methods, specifically, in terms of normalization of observed counts/fluxes and computation of correlation matrices. 

\subsubsection{Normalization in Practice}

Because X-ray data analysis packages such as {\tt xspec} return the $f_{ij}$ and their
uncertainties, it is convenient to rewrite Eq.~\eqref{eq:jasa} in terms of 

\begin{equation}
 \tilde{y}_{ij} 
= \log \frac{f_{ij}}{\tilde f} = \log \frac{c_{ij}}{T_{ij} a_i\tilde{f}},
\quad
\tilde{B_i} = \log \frac{A_i}{a_i},  
\quad \hbox{and} \quad
\tilde{G_j} = \log \frac{F_j}{\tilde{f}} 
\label{eq:transform}
\end{equation}

\noindent
to obtain
\begin{equation}
    \tilde{y}_{ij} = \tilde{B_i} + \tilde{G_j} - \frac{1}{2 \sigma_i^2}  + e_{ij},
\label{eq:solveme}
\end{equation}

\noindent
where $\tilde{f}$ is a fiducial flux (usually the maximum of the $f_{ij}$) used to normalize the data to the range $[0,1]$.
Model~\eqref{eq:solveme} is functionally equivalent to Model~\eqref{eq:jasa}.
This definition of $\tilde{G}_j$, normalized by $\tilde{f}$, which depends on data, is only introduced for computational convenience and does not affect the model or its interpretation. {Technically, using a data-dependent ``parameter", here $\tilde G_j$, implies a data-dependent prior distribution, which is generally not legitimate from a Bayesian viewpoint. However, because using a flat prior on $G_j$ is the same as using a flat prior on $\tilde G_j=G_j-\log{\tilde f}$, there is no actual effect in our implementation.}
Thus, Model~\eqref{eq:solveme} has the same form as Model~\eqref{eq:jasa}, so
we can embed it into the Bayesian hierarchical model described in Paper I to obtain the full posterior distribution, estimates, and error bars for the $\tilde{B}_i$ and $\tilde{G}_j$. Paper I suggests setting $b_i=0$  (because $\tilde{B}_i  = 0$ implies that $A_i = a_i$).

\subsubsection{Deriving Correlations in Practice}\label{sec:MCcalsample}

We proceed by computing numerous instances of instrument effective areas that are varied in controlled ways dictated by current knowledge of uncertainties in calibration. The basis of the method is to generate a so-called calibration sample of areas that represents the range of uncertainties of the effective area, {\em including all the correlations between different energies}. The approach to generating the calibration samples for the different instruments we study here is common to all instruments, with some additional complexity built into the {\it Chandra} samples. We describe the method in brief below and refer the reader to \citet{2006SPIE.6270E..1ID} for more complete description.

We devise a ``perturbation function'' that comprises piecewise cubic segments that stretch between the natural absorption edges of the different materials encountered along the optical path for a given instrument. This function varies about unity by random amounts but is constrained within fixed limits based on specified calibration uncertainties by the cubic function whose parameters are randomly drawn from a truncated Gaussian distribution. The perturbation function is applied as a multiplicative factor to the different subassembly component contributions to the effective area, which are considered on a case-by-case basis.  For instance, in the case of \axaf/ACIS, six plausible mirror effective areas are used, uncertainties in the optical blocking filter and contamination and contamination transmittance are modeled by altering the optical depth of each chemical component within their known uncertainties and recomputing ensemble transmittance, and CCD quantum efficiencies are computed for different realizations of depletion depth and SiO$_2$ layers.  Details of this process are given in \citep[][see also Drake et al.\ 2021, in preparation]{2006SPIE.6270E..1ID}.

\axaf/HETGS and \axaf/LETGS use a similar approach with additional perturbation functions applied for the transmission grating diffraction efficiences. In the case of the other instruments considered here, the perturbation function approach alone is used.

\section{Observations and Data Processing}
\label{sec:data}

Three data sets are considered in Paper I and we add a fourth in this paper.  Here, we detail
how the data sets are handled and the required data processing. 

\subsection{Supernova Remnant 1E0102.2-7219}
\label{sec:e0102}

As in Paper I, the fluxes of the emission line complexes
of O and Ne in the X-ray spectra of SNR~1E0102.2$-$7219
are taken from the detailed comparison of 13 instruments
by \citet{2017A&A...597A..35P}.
Briefly, the spectra of each non-dispersive instrument are
fit with a model with five free parameters: an overall normalization
and four emission line fluxes of O {\sc vii}, O {\sc viii}, Ne {\sc ix} and Ne {\sc x}.
Because the emission lines of the same element have similar energies,
their effective areas are comparable and highly correlated, so we combined
the O and Ne line fluxes to create two fluxes
for each instrument. In our statistical analysis, these two fluxes are treated as  ``sources'': one for O and another for Ne.
The data are normalized to the O or Ne fluxes obtained
by the \xmm\ pn instrument.  By requiring that each instrument
analysis uses the same model, except primarily for the strengths of the
emission lines, the $q(E;\theta_j)$ values do not depend on the
instrument, nor on the measured line fluxes.

Table~\ref{tab:tau_e0102} shows the $\tau_i$ values assigned to the effective areas
for each instrument considered.  The values were taken from Table~\ref{tab:tau1}
using the bandpass that covers the lines of interest: 0.54-0.80 keV for O
and 0.8-1.2 keV for Ne.  For the correlation matrix, there is only one
off-diagonal term, which we set to 0.88 {for ACIS instruments and 0.82 for XMM instruments}, as derived as in
\S~\ref{sec:correlation}.
 
\begin{deluxetable}{c|cc}
  \tablecolumns{5}
  \tablewidth{0pc}
  \tablecaption{Heterogeneous $\tau_i$ Values for 1E0102
    Analysis\tablenotemark{a} \label{tab:tau_e0102} }
  \tablehead{
	\colhead{Instrument} &	\colhead{Oxygen} & \colhead{Neon} }
\startdata
XMM/RGS1 &	5 &5\\
XMM/MOS1 &	6  &6\\
XMM/MOS2 &	6  &6\\
XMM/pn	      &   2 & 2\\
ACIS-S3	     &    3  &3\\
ACIS-I3    &     	3  &3\\
ACIS/HETG &	3  &3\\
Suzaku/XIS0 &	15   &10\\
Suzaku/XIS1 &	15   &10\\
Suzaku/XIS2 &	15   &10\\
Suzaku/XIS3 &	15   &10\\
Swift/XRT-WT &	10  &7.5\\
Swift/XRT-PC &	10  &7.5\\
\enddata
\tablenotetext{a}{Values for $\tau_i$ are in percentages for each combination of instrument
and line complex, using $\tau_i$ values taken from Table~\ref{tab:tau1}.
}
\end{deluxetable}

\subsection{Sources from the 2XMM Catalog}\label{sec:2xmm}

We select a sample of X-ray sources
from the 2$^{nd}$ European Photon Imaging Camera (EPIC)
Serendipitous Source (2XMM) Catalog \citep{2009A&A...493..339W}.
EPIC consists of three X-ray cameras with CCD sensors mounted on the
ESA spacecraft \xmm\ \citep{2001A&A...365L...1J};
the EPIC-pn \citep{2001A&A...365L..18S} and two EPIC-MOS \citep{2001A&A...365L..27T}
cameras observe celestial sources quasi-simultaneously within their co-aligned fields of view.
For this analysis, v14.0 of the Science Analysis
System \citep[SAS;][]{2004ASPC..314..759G} is used, as well as the calibration files as available in 2016.
A description of the data reduction and spectral extraction procedure appears in  \cite{2014A&A...564A..75R}.
Soft, medium, and hard bands are defined to be the 0.5-1.5, 1.5-2.5, and 2.5-10 keV bands,
respectively.
Due to variability,different observations of the same source are treated as separate sources for a total of 41 observations of 35 distinct sources.
The normalizing (maximum) fluxes in the soft, medium,
and hard bands are 0.138, 0.000701, and 0.00223 photons cm$^{-2}$ s$^{-1}$,
the brightest sources in their respective lists.
The data are provided in tables \ref{tab:2xmmsoft}-\ref{tab:2xmmhard} {in the Appendix}.
There are two primary
features of the analysis procedure that make the results suitable
for Concordance analysis: 1) the count spectra
for the different instruments (pn, MOS1, MOS2) are fit simultaneously to
power laws, so
that the spectral slopes, $\theta_j = \Gamma_j$ (where $n_E[\Theta_j] = n_j E^{-\Gamma_j}$)
do not depend on the instrument combination, and 2) the sources are faint
enough that pileup is not an issue.

Table~\ref{tab:tau_xmm} gives the $\tau_i$ values assigned to the effective areas
for the pn and MOS instruments and the three bandpasses.
The values are taken from Table~\ref{tab:tau1}
using the 0.54-0.8 and 0.8-1.2 keV $\tau_i$ values
for the soft band, the 1.2-2.2 keV $\tau_i$ value for the medium
band, and an average of the 2.2-10 keV $\tau_i$
values for the hard band.
Table~\ref{tab:cormat_2xmm} gives the correlation matrix values, $\rho_{mn}$, computed
for the pn and MOS instruments and the three bandpasses used in the 2XMM catalog
while Table~\ref{tab:cormat_xcal} provides these values for the bandpasses
used in the XCAL analysis.
For simplicity, we assume that the $\rho_{mn}$ are the same for each instrument.

\begin{deluxetable}{l|ccc}
  \tablecolumns{5}
  \tablewidth{0pc}
  \tablecaption{Heterogeneous $\tau$ Values for 2XMM and XCAL Analyses\tablenotemark{a} \label{tab:tau_xmm} }
  \tablehead{
	\colhead{Instrument} &	\colhead{Soft band} & \colhead{Medium band} & \colhead{Hard band} }
\startdata
pn &2 &2 &2.3\\
MOS1 &6 &6 &7.3\\
MOS2 &6 &6 &7.3\\
\enddata
\tablenotetext{a}{Values for $\tau$ are percentages for each combination of instrument and line complex, using $\tau$ values from Table~\ref{tab:tau1}.}
\end{deluxetable}

\begin{deluxetable}{l|ccc}
  \tablecolumns{4}
  \tablewidth{0pc}
    \tablecaption{Correlation matrix for 2XMM Analyses \label{tab:cormat_2xmm}}
\tablehead{
    \colhead{Band} & \colhead{Soft band} & \colhead{Medium band} & \colhead{Hard band} }
    \startdata
    Soft band & 1 & 0.61 & 0.13 \\
    Medium band & 0.61 & 1 & 0.53 \\
    Hard band & 0.13 & 0.53 & 1 \\
 \enddata
\end{deluxetable}

\begin{deluxetable}{l|ccc}
  \tablecolumns{4}
  \tablewidth{0pc}
    \tablecaption{Correlation matrix for XCAL Analyses \label{tab:cormat_xcal}}
\tablehead{
    \colhead{Band} & \colhead{Soft band} & \colhead{Medium band} & \colhead{Hard band} }
    \startdata
    Soft band & 1 & 0.63 & 0.20 \\
    Medium band & 0.63 & 1 & 0.52 \\
    Hard band & 0.20 & 0.52 & 1 \\
 \enddata
\end{deluxetable}

\subsection{Active Galaxies from the XCAL Sample}\label{sec:xcal}

Another set of EPIC spectra used to validate the model is the so-called
``\xmm\ Cross-Calibration'' (XCAL) sample.\footnote{Details of the XCAL
processing are available in Section~4 of the XMM calibration
memo {\tt XMM-SOC-CAL-TN-0052}, by Stuhlinger et al.\ (2010).
The memo is available at
{\tt https://xmmweb.esac.esa.int/docs/documents/CAL-TN-0052.ps.gz}.}
This is a sample of radio-loud Active Galactic Nuclei (AGN), primarily blazars,
observed routinely by \xmm\ in the framework of its in-flight calibration
program \citep{2015JATIS...1d7001G}.
As with the 2XMM sample described in \S3.2, the sources are variable.
In this case, there are more blazars and high signal observations, for a total of 108, 103, and 94 observations of 22 distinct sources
in the soft, medium, and hard bands that exceeded a flux limit
criterion without highly discrepant fluxes between the three instruments.
The normalizing (maximum) fluxes in the soft, medium,
and hard bands are set to 0.126, 0.0156, and 0.0154 photons cm$^{-2}$ s$^{-1}$, the brightest sources in their respective lists.
{Data are provided in tables \ref{tab:xcalsoft}-\ref{tab:xcalhard}) in the Appendix}.
As with the 2XMM sources, the pn, MOS1, and MOS2 data were fit simultaneously to
power law spectra so that $\Gamma_j$ is the same for each instrument.
However, compared to the 2XMM sources, the XCAL
sources are bright, often exceeding the count rate threshold beyond which
the fraction
of events affected by pile-up is no longer
negligible \citep{2015A&A...581A.104J}. Spatial
regions on the detector affected by pile-up are removed by excising the
core of the telescope Point Spread Function (PSF) up to an
observation-dependent radius.  This radius is determined on the basis of the ratio between non X-ray diagonal and standard X-ray
``patterns'' (measure of the event shape in the CCD)
in EPIC-MOS \citep{2015A&A...581A.104J}; and by visual
inspection of the pattern distribution curves in
EPIC-pn using the SAS task {\tt epatplot} in EPIC-pn. This ``PSF core
excising method'', while unavoidable to retain the highest possible fidelity
of event spectral calibration, may introduce excess variance in
the $f_{ij}$ via systematic uncertainty in the
energy-dependent correction for the fraction of events scattered into or out of the
annular spectral extraction region by the
PSF (the so-called ``Encircled Energy Correction'' fraction).
Values for $\tau_i$ and $\rho_{mn}$ are the same as for the 2XMM sample.

\subsection{Active Binary Capella}
\label{sec:capella}

Capella ($\alpha$~Aur~AB; G1{\tt{III}}+G8{\tt{III}}; 13pc) is a spectroscopic binary that is the brightest line-dominated source accessible to non-Solar X-ray missions.  It is remarkably steady for a coronal source, having never exhibited significant flaring. While it does vary over timescales of months, it does not show any evidence of flux variability over timescales of weeks or less.  Consequently, it has often been used as a calibration target, in particular with \axaf.  It has been observed several times with different detector and grating combinations in close proximity (see Table~\ref{tab:capella_obs} {in the Appendix}).  These observations allow us to carry out an assessment of the internal cross-calibration of the \axaf\  grating spectrometers.

We estimated the total fluxes in each of several strong lines: the highly-ionized lines of \FeXVII\ (at 15\AA\ and 17\AA) whose formation temperatures overlap the peak emission measures of Capella, and the hydrogenic lines of \NeX\ (12.13\AA) and \OVIII\ (18.96\AA).
For the purposes of this calculation, we treat each of
the four emission lines as different sources.
We then form 21 epoch
groups, comprised of observations that
are within 0.1 yr of each other\footnote{Aug/Sep99,
Mar00, Feb01, Apr02, Oct02, Sep03, Sep04, Mar05,
Oct05, Apr06, Apr07, Apr08, Apr09, Nov09, Nov/Dec10, Dec11, Dec13, Dec14, Jul16, Sep16,
and Dec18; see Table~\ref{tab:capella_obs}.}, giving a total of 84 sources.
Similarly, $+1$ and $-1$ grating orders are treated distinctly for each of
four grating/detector combinations, ACIS-S/HEG, ACIS-S/MEG, ACIS-S/LEG, and HRC-S/LEG,
for a total of eight instruments.
The fluxes were normalized to the maximum values for each emission line:
28.02, 60.23, 49.66, and 23.33 $\times$
$10^{-13}$~erg~s$^{-1}$~cm$^{-2}$ for \OVIII, \FeXVII\ 17\AA, \FeXVII\ 15 \AA, and \NeX, respectively.
We use {\sl CIAO}v4.11 to extract the dispersed spectra, and compute the effective areas using the contamination corrections as in {\tt CALDBv4.8.0.1}.

The values of $\tau_i$ are taken from Table~\ref{tab:tau_capella} and the
correlation matrix is given in Table~\ref{tab:cormat_capella}.

\section{Results}\label{sec:results}

Here we present results from new measurements and extensions to the results in Paper I.
In each case, we generate 10,000 Monte Carlo replicates from the respective posterior distributions as the basis for our statistical inferences.

\subsection{1E0102}
\label{sec:1e0102}

These data provide a illustration of a case where there are many instruments that obtain
data on the same source, shown in Figs.~\ref{fig:oxygen} and \ref{fig:neon}.
The effect of allowing heterogeneous $\tau$ values is apparent in both cases.
Generally,
when the prior distribution on an instrument's effective area is more uncertain than average,
giving a relatively large value of $\tau_i$, then the data for that instrument
is given less weight, so the posterior estimate of its effective area
is more likely to deviate from the prior estimate by comparison to when
all instruments have equally uncertain prior estimates.  In addition, the posterior range of the deviation
is more likely to be large when $\tau_i$ is larger.  The \suzaku\ results
for the O lines all show this effect.
When effective areas are correlated between the O and Ne data sets, the ACIS-I3
point is particularly affected due to the discrepant results obtained when Ne and
and O data are considered independently.

\begin{figure}[tbph]
    \centering
    \includegraphics[width=\textwidth]{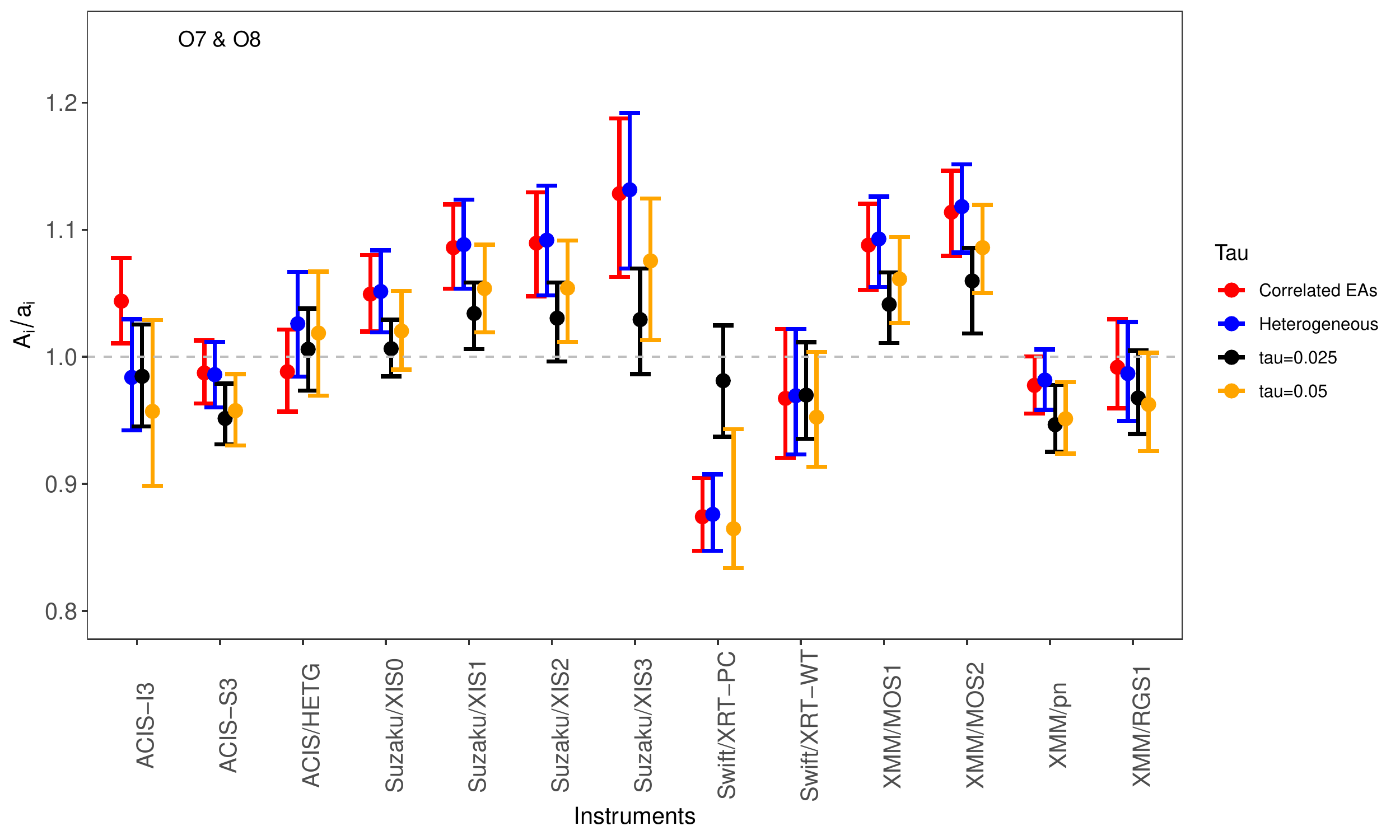}
    \caption{Results of the Concordance analysis for the data from the SNR
    1E0102 for the combination of fluxes of the lines of O {\sc vii} and O {\sc viii}.
    The $\tau = 0.025$ (in black) and $0.05$ (in yellow) results are the
    same as given in Paper I and are shown to elucidate the effects of including
    heterogeneous $\tau$ values (in blue) and adding effective area correlations
    (in red).  The error bars represent the 90\% (5\%-95\%) confidence regions on the
    posterior estimate of $ A_i/a_i $, as defined in \S\ref{sec:method}.
    When effective areas are correlated between the O and Ne data sets, the ACIS-I3
    point is particularly affected due to the discrepant results obtained when Ne and
    and O data are considered independently.}
    \label{fig:oxygen}
\end{figure}

\begin{figure}[tbph]
    \centering
    \includegraphics[width=\textwidth]{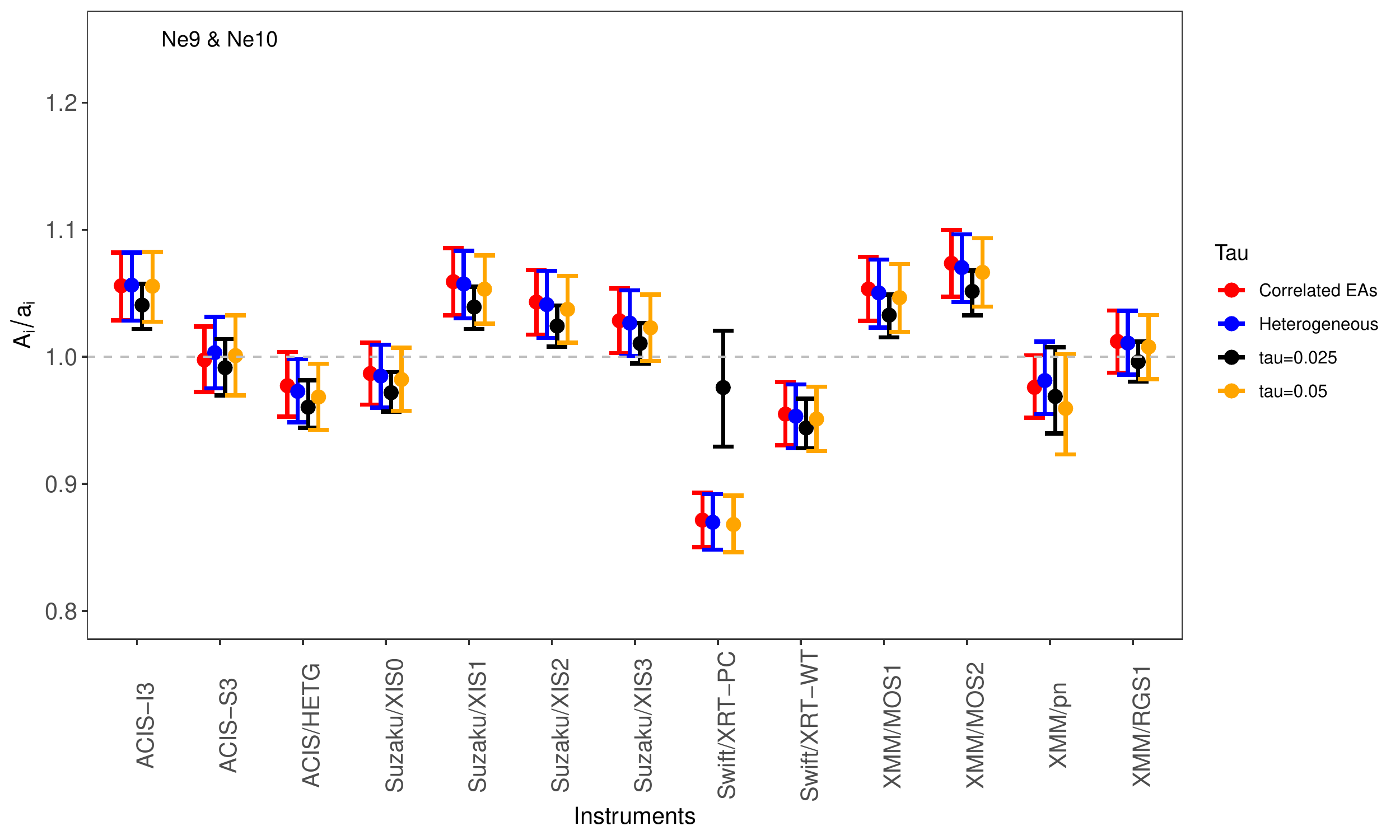}
    \caption{Same as Fig.~\ref{fig:oxygen} except for the combination of fluxes of the lines of
    Ne {\sc ix} and Ne {\sc x} (see text).
    Unlike the case for the O lines, the posterior estimates of the effective
    area tend to be more stable in this band and relatively independent of the
    uncertainties in the effective area priors.}
    \label{fig:neon}
\end{figure}

\subsection{XMM Samples}

Figures~\ref{fig:2xmm} and \ref{fig:xcal} show the results of the Concordance analysis
for the two \xmm\ data.  In this example, there are many sources and few instruments,
in contrast with the 1E0102 data set.
The 2XMM results show a high degree of consistency between the instruments, consistently
favoring 3-5\% increases to the effective areas of the MOS detectors across all bands
and a corresponding, slight decrease to the pn effective area.  With the use of
individualized $\tau$ values, the Concordance analysis drives the pn effective areas
toward the prior, as one might expect due to the significantly smaller $\tau$ assigned
to the pn compared to the MOS detectors.

The XCAL sample, shows similar trends to that of the 2XMM sample with with a more significant indication that the MOS2 detector's effective area should be increased
2-3\% more than that of the MOS1 detector.

\begin{figure}[tbph]
    \centering
    \includegraphics[width=\textwidth]{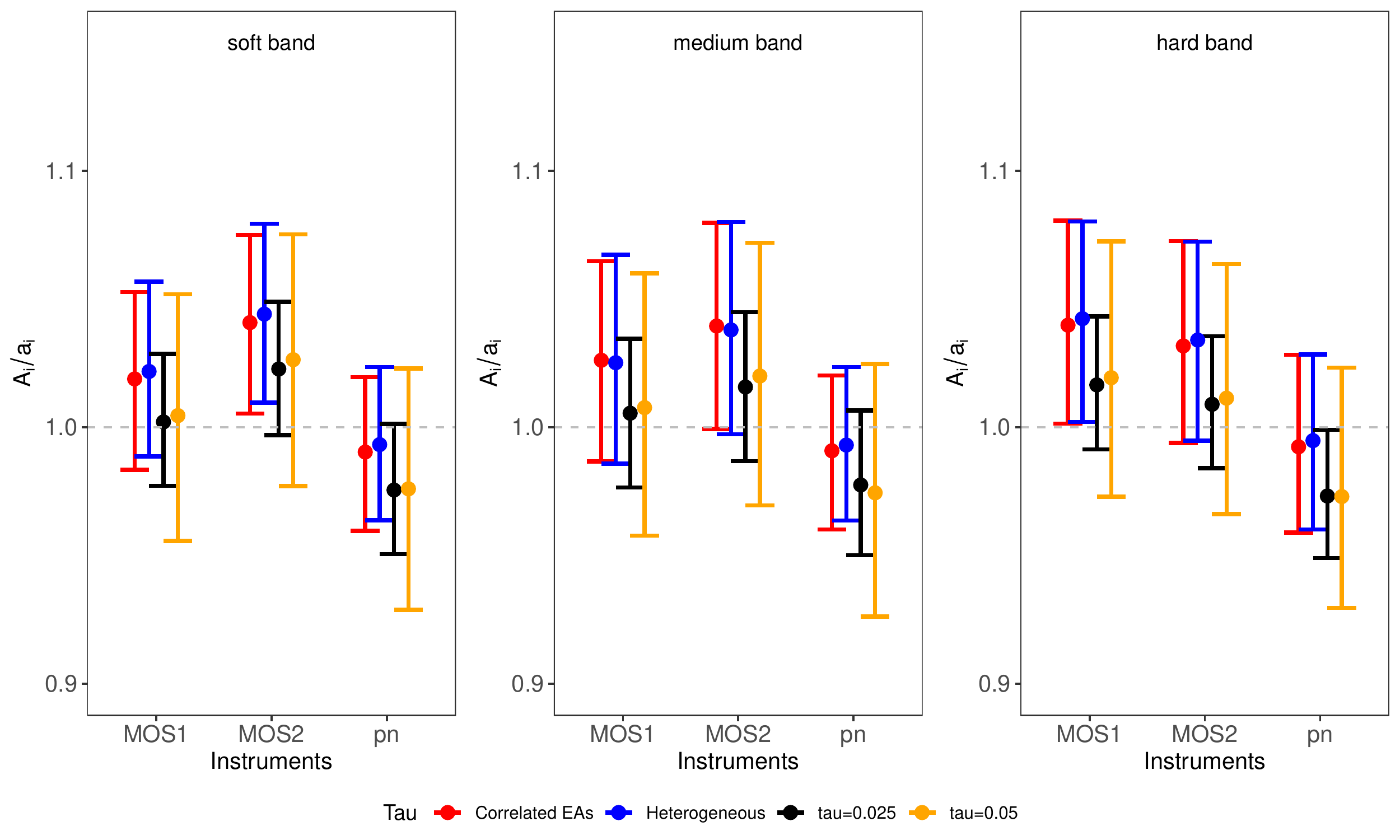}
    \caption{Concordance results for the 2XMM sample.  Results are color-coded
    as in Fig.~\ref{fig:oxygen}.  When the $\tau$ values are allowed to vary
    by instrument, the ``heterogeneous'' case, the posterior for the pn
    centers on the prior, due to the smaller value of $\tau$ than used for
    the MOS detectors.  At the same time, higher effective areas for the MOS
    detectors are indicated across all bands.}
    \label{fig:2xmm}
\end{figure}

\begin{figure}[tbph]
    \centering
    \includegraphics[width=\textwidth]{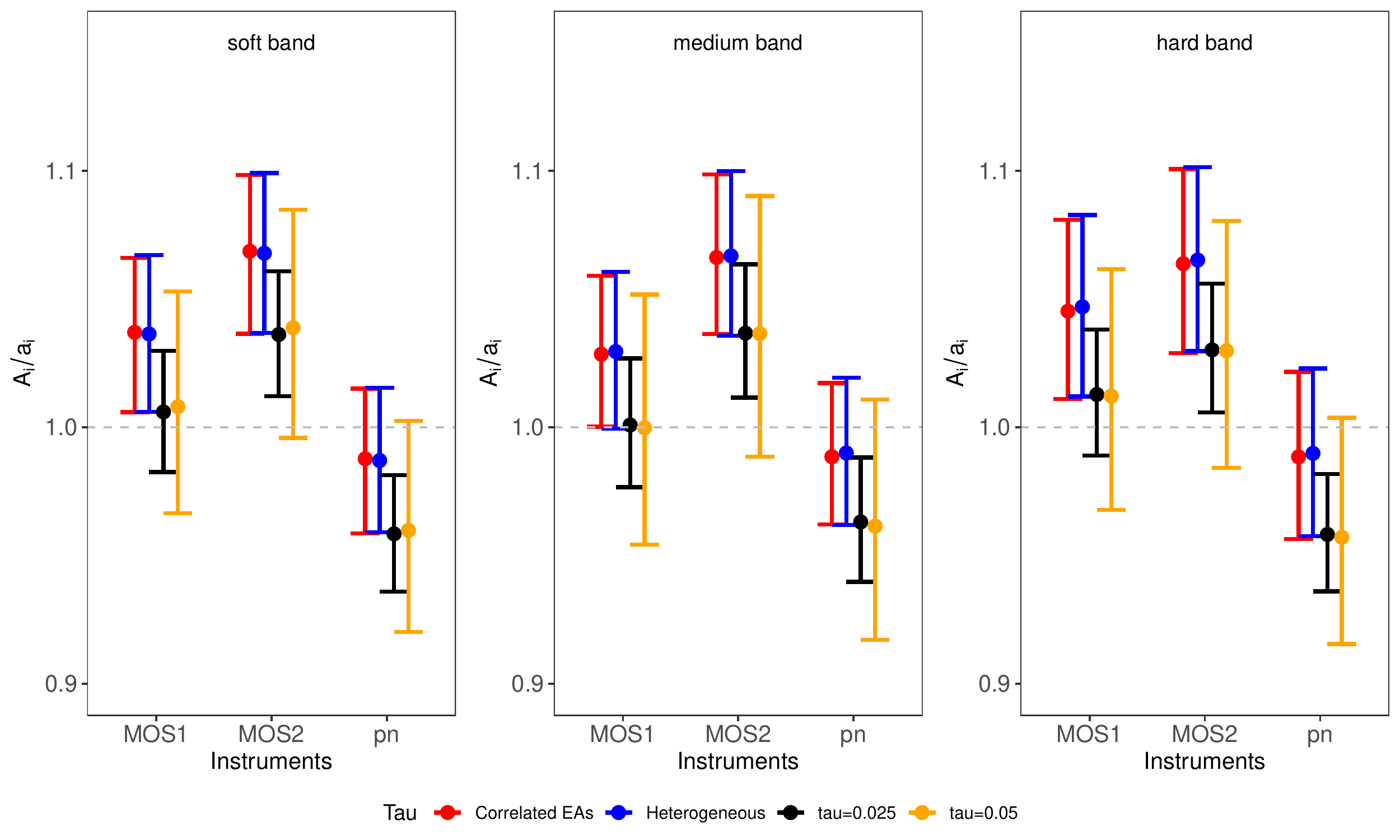}
    \caption{Same as Fig.~\ref{fig:2xmm} except for the XCAL sample.  These
    results are generally consistent with those from the 2XMM sample but with
    a somewhat stronger indication that the MOS2 effective area should be
    increased relative to MOS1}
    \label{fig:xcal}
\end{figure}

\subsection{Capella Line Fluxes for \axaf\ Grating Spectrometers}
\label{sec:capellaresults}

Results from the Concordance analysis as applied to the Capella data are
shown in Fig.~\ref{fig:capella}.  There are several features of interest.
First, the effective area
corrections for the LETGS (ALEG and HLEG) are generally negative while
those of the HETGS (HEG and MEG) are generally positive.  These
corrections are consistent with preliminary results on independent
data where the instruments are cross-calibrated with alternating
observations of Mk 421.
Second, the $+1$ and $-1$ orders generally agree well for all instruments
and wavelengths.
Third, when the effective area correlations are included, the posterior
effective areas for the longer wavelengths (\OVIII\ and \FeXVII\ $\lambda 17$)
more strongly deviate from their priors.

\begin{figure}[tbph]
    \centering
    \includegraphics[width=0.75\textwidth]{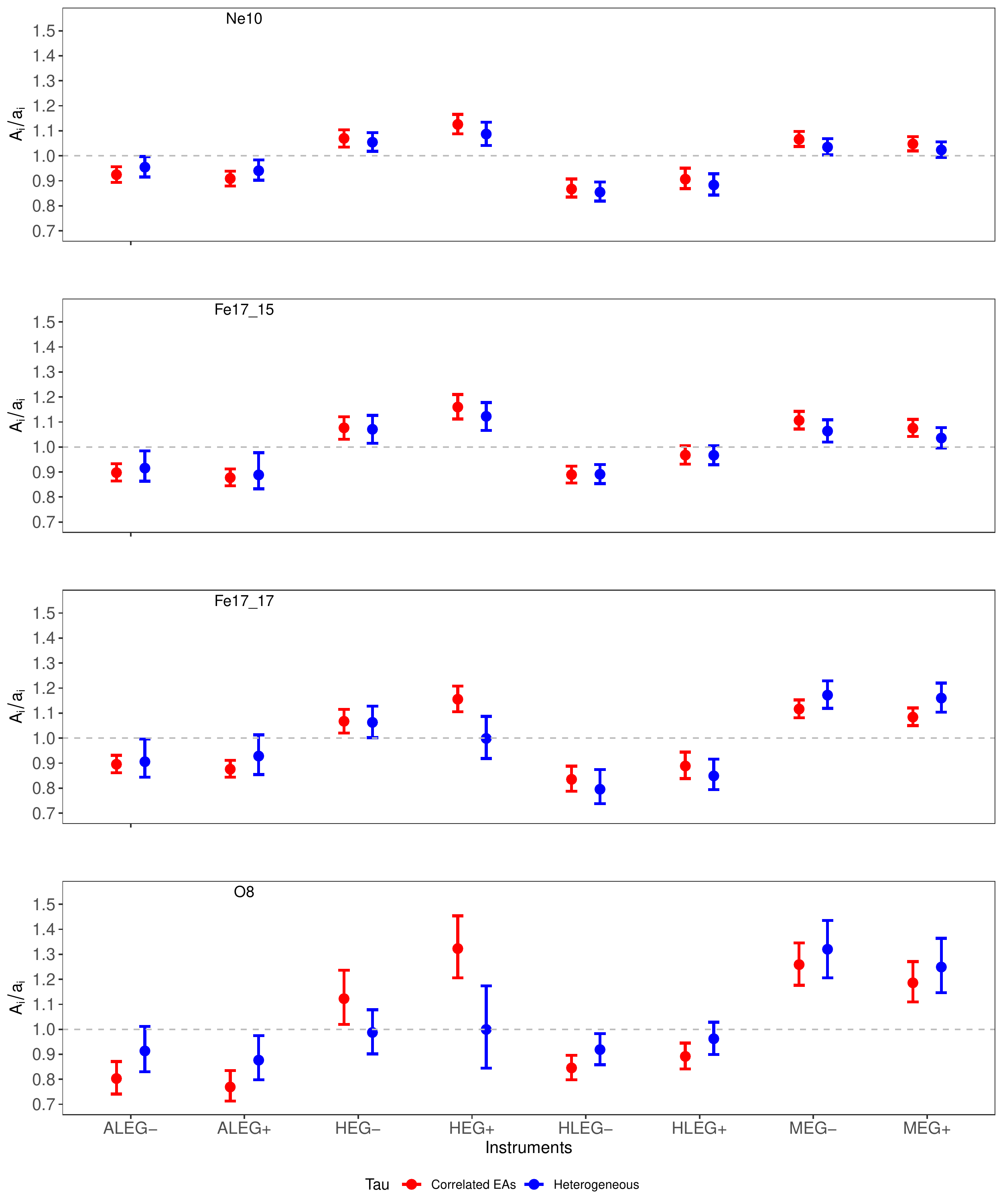}
    \caption{Concordance analysis using measurements of four emission lines
    using the the \axaf\ grating spectrometers. Emission lines used in
    the various panels are: a) \NeX, b) \FeXVII\ $\lambda 15$,
    c) \FeXVII\ $\lambda 17$, d) \OVIII.  Color coding of results are as in
    Fig.~\ref{fig:oxygen}.  Features to note are 1) that the effective area
    corrections for the LETGS (ALEG and HLEG) are generally negative while
    those of the HETGS (HEG and MEG) are generally positive and 2) the
    $+1$ and $-1$ orders generally agree well.}
    \label{fig:capella}
\end{figure}

\subsection{Method Validation and Assessment}

Paper I includes a series of numerical studies that explore the statistical properties of our method. For example, Figure~2 of Paper I illustrates that our posterior distributions of the effective areas
cover the true effective areas in a simulation study.
Figure 7 of Paper I then goes on to contrast the estimated 95\% intervals for log-fluxes constructed using the standard instrument-specific estimates with the combined estimate based on our posterior distribution, illustrating how our Bayesian process achieves a single consistent estimate for each flux but with smaller errors than the standard estimates. Here
we consider additional ways to evaluate how robust our method's results may be, supplementing the simulations performed in Paper I.

\subsubsection{Simulation Studies}

\label{sec:posteriors}
The method developed and applied in Paper I produces Bayesian
posterior distributions for each estimated quantity.
The main quantities of interest here are the fractional corrections to
instrument effective areas, given by $\log A_i - \log a_i$, and the
fractional corrections to the estimated fluxes of sources, given by
$\log F_j - \log f_{ij}$.

We demonstrate how the Concordance method
yields accurate and reliable estimates of $A_i$ and $F_j$ with a simulation study.
The simulation involves 40 simulated sources observed by each of five instruments.
We set the prior means of
the effective areas of the instruments to differ from their actual effective areas
by $\log A_i - \log a_i = [0, 1, -1, 2, -2]$, for $i = 1..5$, respectively.
For example, for instrument 4, the true effective area is systematically
higher than the prior mean
by a factor of $e^2$, resulting
in flux estimates that are systematically too high compared to the true
values.  The $\tau_i$ values are all 1.0 in this simulation, indicating
large uncertainties in prior estimates of the effective areas, and the
measurement uncertainties (i.e., $\sigma_{ij}$) are all set to 0.5 on the log scale, similarly
indicating large uncertainties, except for
instrument 5, for which $\sigma = 0.1$.
This simulation setup is designed to test the robustness of the Concordance
method to data from an instrument with high signal/noise but a systematically
biased effective area.
We replicated this entire set up 200 times and processed each replicate with our Concordance method. A representative replicate 
is shown in Fig.~\ref{fig:simresults}.
The simulations demonstrate
that the Concordance method provides source flux
estimates that are substantially better than would be obtained by simply using the
prior means as estimates of the effective areas.  The replicate simulations
indicate
that the 95\% equal-tailed posterior intervals cover the true values of the effective areas
and source fluxes over 99\% of the time. 
Thus, we not only obtain
better estimates of the source fluxes, but also estimate the effective area corrections well.
Note that the instrument-specific flux estimates
can deviate substantially from the true values for any given source, so that when using a set of
such estimates, the {\it weighted average ratio estimators} for the flux 
(i.e., $(\sum_j f_{ij}/\sigma_j^2 ) / \sum_j 1/\sigma_j^2$)
would be generally biased toward the instrument
with the highest signal/noise, regardless of the accuracy of the instrument's effective area.

\begin{figure}[t!]
\includegraphics[width=17cm]{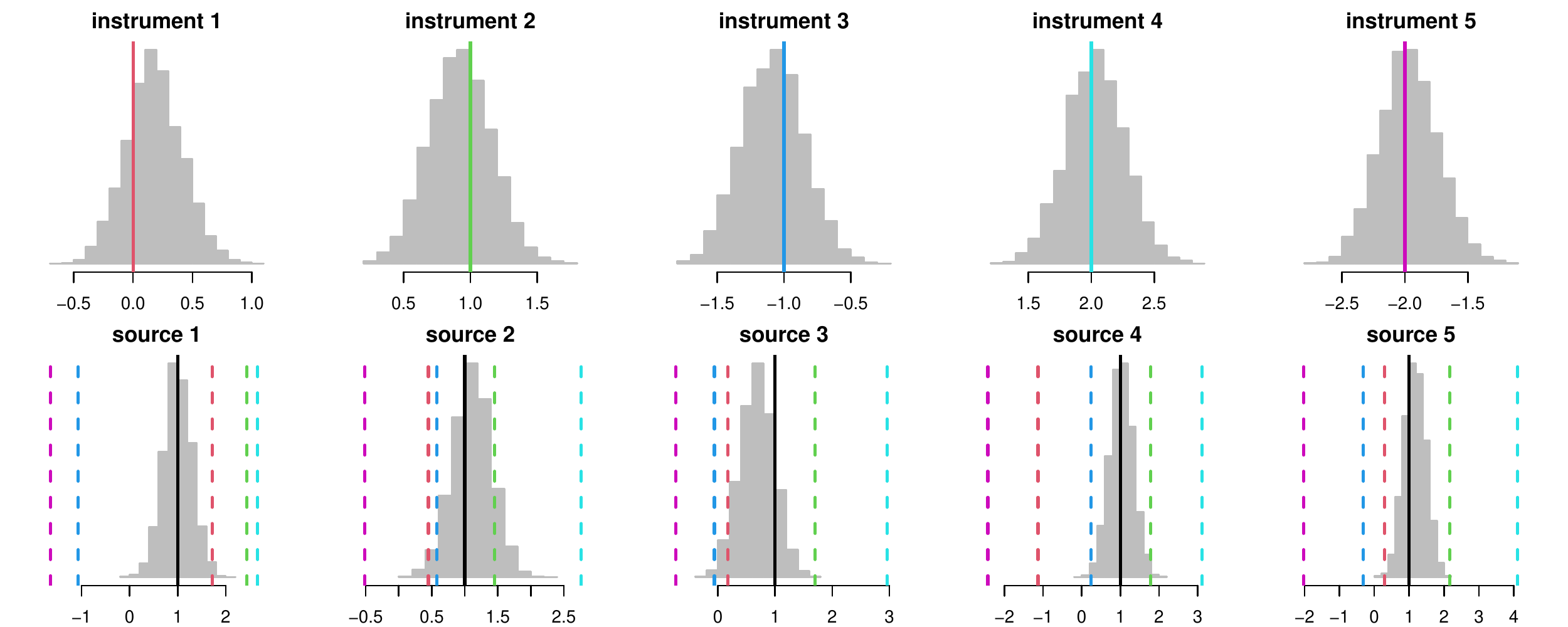}
\caption{ Posterior distributions of the logarithms of the effective area corrections
(top row) and on the logarithms of the source fluxes (bottom row, for five sources) for one of 200 replicate simulations, each involving
five instruments and 40 sources.
In the simulation, all sources have
a true value of $\log F_j = 1$ but the prior means of the instrument effective areas are off
by factors of $\exp( [0,1,-1,2,-2] )$ for $i=1,\ldots,5$.
The top row shows that the Concordance method generates posterior distributions
of the effective area corrections that are well centered on the
true values for each instrument (shown as vertical solid lines in various colors).
The bottom row shows that the posterior distributions of the source fluxes
are well centered on the true values (vertical solid black lines), even while the instrument-specific estimates based on the
prior means of the effective areas (vertical dotted lines of colors corresponding to those of
the instruments in the top row) can be individually erroneous by large factors.
\label{fig:simresults} }
\end{figure}

With an additional pair of simulations, we also quantified the improvement that can be obtained with the Concordance method.  In the first case, we simulated $M=3$ instruments with $A_i/a_i = [1,1,0.9]$ for $i=[1,2,3]$, $\tau_i = 0.05$, and $N=20$ sources with true fluxes all equal to 1.  There were 200 independent simulations and analyses for each setup.  The sources were assumed to have good signal/noise as might be expected for calibration observations: $\sigma_i = 0.03$ (i.e., the $c_{ij}$ were drawn from a Poisson distribution with a mean of about 1100). The second case is the same as the first except there is a higher statistical precision for observations with instrument 3: $\sigma_3 = 0.003$.  Source fluxes were estimated for each simulation using the Concordance method and also using the 
above-mentioned weighted average of ratio estimators:
$(\sum_j f_{ij}/\sigma_j^2 ) / \sum_j 1/\sigma_j^2$.  The 95\% uncertainty bounds on the flux estimates and the coverage fraction where the true flux is included within the uncertainty bounds are shown in Table~\ref{tab:simresults}.  The Concordance estimator generated uncertainty intervals that were accurate and covered the ground truth, in contrast to the ratio estimator, which, despite the widths of the confidence intervals being nominally smaller, generated biased estimates and did not cover the true fluxes.  The situation was worse for the second case, where the Concordance intervals did not change appreciably, but the ratio estimators were biased low by $\approx$10\%, reflecting the higher statistical weight given to an instrument with a biased estimate of its effective area.  Indeed, in this latter case, there was only one source (of 20) in only one simulation (of 200) where the ratio estimator confidence interval included the true value.  This pair of simulation setups illustrates the robustness of the Concordance method to erroneous effective area priors, showing that it is appropriate to use in calibration work where robustness and accuracy are highly valued.

\begin{deluxetable}{c|cccc}   
  \tablecolumns{5}
  \tablewidth{0pc}
  \tablecaption{Results from Two Concordance Simulations \label{tab:simresults} }
    \tablehead{
    \colhead{Simulation Setup\tablenotemark{a}} & \colhead{Flux Estimation} & \multicolumn{2}{c}{95\% Flux Range\tablenotemark{b}} & \colhead{ $r_{95}$\tablenotemark{c}} \\
	\colhead{} &	\colhead{Method} & \colhead{$F_{\rm lo}$} & \colhead{$F_{\rm hi}$}  &	\colhead{}}
\startdata
1 &	Concordance &	0.903   &	1.033   &	0.964 \\
1 &	Ratio Estimator &	0.927 &	0.994 &	0.372 \\
2 &	Concordance &	0.905 &	1.031 &	0.990 \\
2 &	Ratio Estimator &	0.896 &	0.907 &	0.000 \\
\enddata
\tablenotetext{a}{Setups 1 and 2 are the same except that instrument 3 (of 3)
has 3\% statistical
errors for setup 1 and 0.3\% statistical errors for setup 2.  The prior for
the effective area of
instrument 3 is 10\% higher than its true value in both setups.
See Section~\ref{sec:posteriors} for details.}
\tablenotetext{b}{Average 95\% confidence intervals for source flux estimates; the true fluxes
for all sources are set to 1.}
\tablenotetext{c}{Fraction of flux estimates covering true fluxes at the 95\% confidence level
out of 200 simulations of 20 sources each.}
\end{deluxetable}

\subsubsection{Posterior Histograms}

We have found that the posterior distributions of the effective area
corrections are typically well described by Gaussians, as shown in
Fig.~\ref{fig:example_histogram}, parts A-C.  These three examples
were randomly chosen among the dozens of such histograms generated in
our analysis of the data from \S\ref{sec:1e0102}-\ref{sec:capellaresults}.
Occasionally, however, there are histograms that are not obviously
Gaussian, so we also show three ``bad'' examples.
In one case,
Fig.~\ref{fig:example_histogram}, part D, there is a distinct ``notch''
in a side of the distribution and in two cases
(Fig.~\ref{fig:example_histogram}, parts E and F), there is noticeable skew --
tails to large fractional corrections.
These three cases were quite rare but give warning that there may
be inconsistencies in the underlying data.  One known source of
error that is not accounted for in our analysis is in the shape
of the response function, $\Phi_k(E)$.  For instruments like ACIS and the
EPIC detectors, the low energy response is somewhat uncertain and
difficult to calibrate.

\begin{figure}
    \centering
    \includegraphics[width=\textwidth]{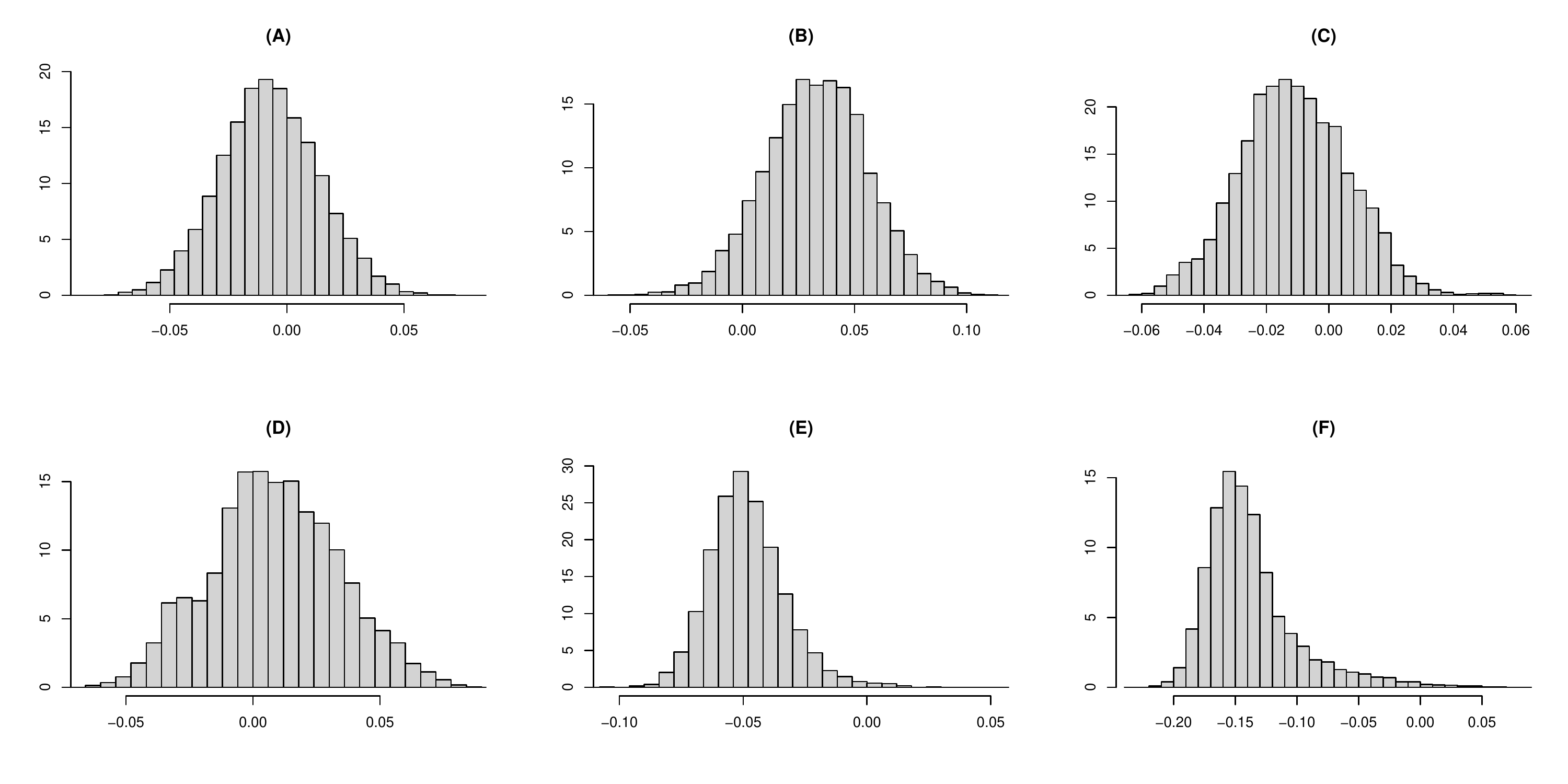}
    \caption{Posterior histograms for the fractional variation of the effective
    area.  The data sets are
    (A) 2XMM data, hard band, XMM/pn, correlated $\tau$ values;
    (B) 2XMM data, hard band, XMM/MOS2, heterogeneous $\tau$ values;
   (C) XCAL data, medium band, XMM/pn, correlated $\tau$ values;
   (D) XCAL data, soft band, XMM/MOS1, $\tau = 0.05$ for all instruments;
   (E) 1E0102 data, O lines, Chandra/ACIS-S3, $\tau = 0.025$ for all instruments; and
   (F) 1E0102 data, O lines, Swift XRT/PC, $\tau = 0.05$ for all instruments.
   Histograms A-C are typical, chosen randomly from several dozen; the distributions
   are well approximated as Gaussians.
   Histograms D-F are atypical, showing skew or other non-Gaussian shapes.
   }
    \label{fig:example_histogram}
\end{figure}

\subsubsection{Sensitivity to Uncertainties in Priors}

The specification of priors is typically under scrutiny for Bayesian analysis in practice. Typically researchers conduct sensitivity analysis to study the outcome sensitivity with respect to small perturbations of the priors. In our setting, the sensitivity of the results with respect to $\tau$ values are revealed by the comparison between heterogeneous $\tau$ values versus the two homogeneous $\tau$ value choices. However, for the correlation matrix in the prior distribution, we adopted Monte Carlo estimates, which is subject to random variations. Thus, we undertook a simple example of the test of sensitivity in this paper, but it is feasible for any user of the concordance tools. Namely, for the Capella data, instead of adopting the full correlation matrix as given in Tables~\ref{tab:cormat_capella} and~\ref{tab:cormat_capella_hrc}, we only keep the correlations between the two Fe bands. Again, we can test out different variations of the correlation matrix with the same procedure and similar analysis. Thorough sensitivity analysis requires extensive testing on a carefully designed set of variations of the prior distributions. See Fig.~\ref{fig:capella_variant} for the results of applying
this variation to the Capella data. By comparing the results of Fig.~\ref{fig:capella_variant} with the original, Fig.~\ref{fig:capella}, we can reveal the sensitivity of the results as opposed to variations of correlations between Fe bands and others, and the correlations of others (Ne and O) within their own. We can see that in Fig.~\ref{fig:capella_variant}, the O and Ne shows nicely aligned results under correlated effective areas versus uncorrelated effective areas. But this is not true for Fig.~\ref{fig:capella}, where O and Ne are still correlated with each other and with other bands, especially for HEG+. Furthermore, while the adjustments for the two Fe chanels are similar across the two figures, the adjustments for Ne are very different across the two figures. The resulting adjustments of effective areas not only deviate more significantly from zero but also have smaller error bars when correlations are taken into account. This makes intuitive sense because the benefit of accounting for correlations among effective areas is to obtain sharper or more informative estimates. 

\begin{figure}
    \centering
    \includegraphics[width=0.75\textwidth]{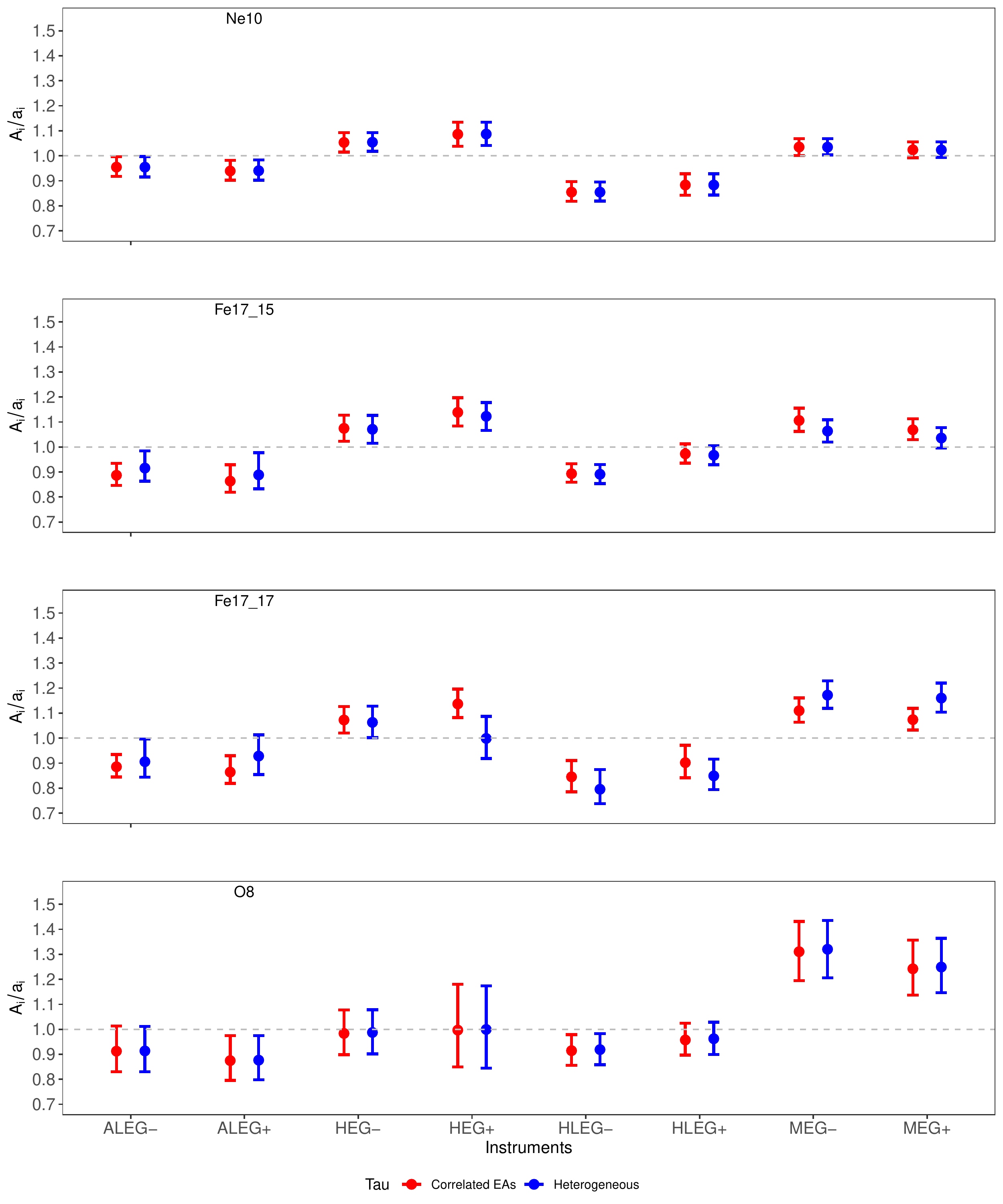}
    \caption{Same as Fig.~\ref{fig:capella} except that only the correlations between the two Fe bands are kept while others are set to be zero, for both ACIS and HRC instruments. }
    \label{fig:capella_variant}
\end{figure}

\section{Conclusion and Directions for Development}\label{sec:summary}

These data sets provided an excellent foundation for the Concordance
project, whose goal is to determine quantitative and objective evidence
for making effective area adjustments in order to improve agreement
between instrument measurements.  The process applied here is
available for use in studies such as we have undertaken.  There are
some avenues to explore for expanding this particular implementation
of the Concordance analysis.

\subsection{Correlations between source bandpass fluxes}

Several types of calibration sources have simple spectra, which is why
they are often used in cross-calibration.  Examples are isolated neutron
stars with blackbody spectra, blazars with power law spectra, and
supernova remnants and clusters of galaxies with thermal spectra.
To the extent that these spectra can be characterized by only a few
parameters, such as a power law slope, then the flux in one band
is closely related to that in an adjacent band.  Furthermore,
many types of source have smoothly continuous spectra -- their
spectral fluxes are tightly correlated on small scales.
Modeling a many bandpasses of a blazar spectrum with a series of
power laws with different slopes would lead to unphysical discontinuities
at bandpass boundaries.
Thus, it would be advantageous to take advantage of this astrophysical
knowledge and include spectral band correlations due to spectral
continuity and simplicity.
The \xmm-\axaf\ blazar XCAL sample is an excellent data set
to examine next, involving three \axaf\ configurations and all four \xmm\ X-ray
detectors and covering the energy range from 0.1 to 10 keV using simultaneous
observations of active galaxies obtained over 20 years of operation.
Preliminary results have been reported at various IACHEC meetings.

\subsection{Secular variations of instrument responses}

While many sources may well vary erratically, instrument behavior can
often be subject to gradual degradation.
With adequate modeling of many observations, one avenue to explore
is how to link instrument effective areas over time within the
Concordance framework.

\subsection{Nonlocal instrument responses}

There are definite difficulties that are encountered when the
detector energy response $\Phi_k(E)$ has a non-Gaussian component,
a broad asymmetry, or bimodality because systematic errors in the
response function can appear in an apparently unrelated bandpass.
Response function errors may be responsible for some of the artifacts in
our posterior histograms (see \S\ref{sec:posteriors}).
The response functions of solid state detectors have escape peaks
that can generate events at a significantly different apparent
energy than that of the incoming photon.  Modeling the effects
of systematic errors in response functions is possible in principle,
especially with methods such as used to determine the effective
correlation function (see \S~\ref{sec:correlation}).  One approach
for dealing with this issue would be to expand the
Concordance mathematical model to include a term to account for variance of
the $T_{ij}$ values.

\acknowledgments

Support for this work was provided in part by the National Aeronautics and
Space Administration (NASA) through the Smithsonian Astrophysical Observatory (SAO)
contract SV3-73016 to MIT for support of the Chandra X-Ray Center (CXC),
which is operated by SAO for and on behalf of NASA under contract NAS8-03060.
Support was also provided by NASA under contract NAS8-39073 to SAO. 
This work was conducted in collaboration with the CHASC International Astrostatistics Center. CHASC is supported by NSF DMS-18-11308, DMS-18-11083, and DMS-18-11661.  DvD was supported in part by a Marie-Skodowska-Curie RISE (H2020-MSCA-RISE-2015-691164, H2020-MSCA-RISE-2019-873089) Grants provided by the European Commission.  

\vspace{5mm}
\facilities{CXO(ACIS), CXO(HRC), CXO(HETGS), CXO(LETGS), XMM-Newton(pn), XMM-Newton(MOS)}

\software{Concordance \url{https://github.com/astrostat/Concordance}, ciao \citep{2006SPIE.6270E..1VF}, SAS \citep{2004ASPC..314..759G}, MCCal \citep{2006SPIE.6270E..1ID},
PINTofALE \citep{2000BASI...28..475K}}.

\clearpage

\begin{deluxetable}{c|cccc}   
  \tablecolumns{5}
  \tablewidth{0pc}
  \tablecaption{Values of $\tau_i$ for Capella Analyses\tablenotemark{a} \label{tab:tau_capella} }
    \tablehead{
	\colhead{Line} &	\colhead{\NeX~$\lambda$12} & \colhead{\FeXVII~$\lambda$15} & \colhead{\FeXVII~$\lambda$17}  &	\colhead{ \OVIII~$\lambda$19}}
\startdata
HEG&	4&	5&	5&	10\\
MEG&	4&	5&	5&	10\\
LEG/A&	4&	5&	5&	10\\
LEG/H&	7&	7&	7&	7\\
\enddata
\tablenotetext{a}{Values for $\tau$ are in percentages.}
\end{deluxetable}

\begin{deluxetable}{c|cccc}
   \tablecolumns{5}
  \tablewidth{0pc}
    \tablecaption{ACIS Correlation matrix used for Capella \label{tab:cormat_capella}}
   \tablehead{
	\colhead{Line} &	\colhead{\NeX~$\lambda$12} & \colhead{\FeXVII~$\lambda$15} & \colhead{\FeXVII~$\lambda$17}  &	\colhead{ \OVIII~$\lambda$19}}
    \startdata
    \NeX~$\lambda$12 & 1 & 0.96 & 0.92 & 0.89 \\
    \FeXVII~$\lambda$15 & 0.96 & 1 & 0.99 & 0.97 \\
    \FeXVII~$\lambda$17 & 0.92 & 0.99 & 1 & 0.99 \\
    \OVIII~$\lambda$19 & 0.89 & 0.97 & 0.99 & 1 \\
    \enddata   
\end{deluxetable}

\begin{deluxetable}{c|cccc}
   \tablecolumns{5}
  \tablewidth{0pc}
    \tablecaption{HRC Correlation matrix used for Capella \label{tab:cormat_capella_hrc}}
   \tablehead{
	\colhead{Line} &	\colhead{\NeX~$\lambda$12} & \colhead{\FeXVII~$\lambda$15} & \colhead{\FeXVII~$\lambda$17}  &	\colhead{ \OVIII~$\lambda$19}}
    \startdata
    \NeX~$\lambda$12 & 1 & 0.84 & 0.71 & 0.62 \\
    \FeXVII~$\lambda$15 & 0.84 & 1 & 0.83 & 0.74 \\
    \FeXVII~$\lambda$17 & 0.71 & 0.83 & 1 & 0.91 \\
    \OVIII~$\lambda$19 & 0.62 & 0.74 & 0.91 & 1 \\
    \enddata   
\end{deluxetable}

\appendix
\section{Measured fluxes}\label{sec:fluxes}

Here we list all the fluxes used in the calculations described above: for sources in the {\em 2XMM} catalog (see Section~\ref{sec:2xmm}) in the soft (Table~\ref{tab:2xmmsoft}), medium (Table~\ref{tab:2xmmmedium}), and hard (Table~\ref{tab:2xmmhard}) bands; for active galaxies from the {\em XCAL} sample (Section~\ref{sec:xcal}) in the soft (Table~\ref{tab:xcalsoft}), medium (Table~\ref{tab:xcalmedium}), and hard (Table~\ref{tab:xcalhard}) bands; and the line fluxes measured during the various Capella grating observations with \axaf\ (Section~\ref{sec:capella}; Table~\ref{tab:capella_obs}).
Note that the data used for the analysis of SNR 1E0102.2-7219 (Section~\ref{sec:e0102}) are given in \citet{concordancejasa}.

\setcounter{table}{0}
\renewcommand\thetable{\Alph{section}.\arabic{table}}

\startlongtable
\begin{deluxetable}{l|cc|cc|cc}
\tablecolumns{7}
\tablewidth{0pc}
\tablecaption{{\em 2XMM} Concordance Fluxes -- Soft Band\tablenotemark{a} \label{tab:2xmmsoft} }
\tablehead{
\colhead{Target} & \multicolumn{2}{c}{pn} & \multicolumn{2}{c}{MOS1} & \multicolumn{2}{c}{MOS2} \\
\colhead{} & \colhead{$f_{ij}$} & \colhead{$\sigma_{ij}$} & \colhead{$f_{ij}$} & \colhead{$\sigma_{ij}$} & \colhead{$f_{ij}$} & \colhead{$\sigma_{ij}$} }
\startdata
            1127$-$145 & 0.0075 & 0.0009 & 0.0074 & 0.0009 & 0.0077 & 0.0010 \\ 
          1E0919$+$515 & 0.0068 & 0.0013 & 0.0081 & 0.0015 & 0.0083 & 0.0015 \\ 
             4C06.41 & 0.0024 & 0.0001 & 0.0026 & 0.0001 & 0.0026 & 0.0001 \\ 
       APM 08279$+$5255 & 0.0035 & 0.0006 & 0.0035 & 0.0006 & 0.0036 & 0.0006 \\ 
               CAL 87 & 0.0125 & 0.0011 & 0.0136 & 0.0012 & 0.0134 & 0.0012 \\ 
              Cen X-4 & 0.0180 & 0.0027 & 0.0180 & 0.0027 & 0.0190 & 0.0028 \\ 
         CoD-33 7795 & 0.1523 & 0.0222 & 0.1578 & 0.0230 & 0.1618 & 0.0236 \\ 
         ESO323$-$G077 & 0.0011 & 0.0004 & 0.0012 & 0.0005 & 0.0012 & 0.0005 \\
\enddata
\tablenotetext{a}{Fluxes are normalized to 0.138 photons cm$^{-2}$ s$^{-1}$.}
\end{deluxetable}

\startlongtable
\begin{deluxetable}{l|cc|cc|cc}
\tablecolumns{7}
\tablewidth{0pc}
\tablecaption{{\em 2XMM} Concordance Fluxes -- Medium Band\tablenotemark{a} \label{tab:2xmmmedium} }
\tablehead{
\colhead{Target} & \multicolumn{2}{c}{pn} & \multicolumn{2}{c}{MOS1} & \multicolumn{2}{c}{MOS2} \\
\colhead{} & \colhead{$f_{ij}$} & \colhead{$\sigma_{ij}$} & \colhead{$f_{ij}$} & \colhead{$\sigma_{ij}$} & \colhead{$f_{ij}$} & \colhead{$\sigma_{ij}$} }
\startdata
            1127-145 & 0.481 & 0.049 & 0.496 & 0.053 & 0.490 & 0.052 \\ 
          1E0919+515 & 0.053 & 0.053 & 0.069 & 0.066 & 0.068 & 0.065 \\ 
             4C06.41 & 0.131 & 0.015 & 0.142 & 0.017 & 0.143 & 0.018 \\ 
       APM08279+5255 & 0.085 & 0.041 & 0.088 & 0.042 & 0.082 & 0.040 \\ 
              CenX-4 & 0.088 & 0.035 & 0.089 & 0.022 & 0.091 & 0.023 \\ 
         CoD-33 7795 & 0.275 & 0.136 & 0.287 & 0.143 & 0.276 & 0.136 \\ 
         ESO323-G077 & 0.425 & 0.184 & 0.438 & 0.202 & 0.439 & 0.203 \\ 
           GRB080411 & 0.348 & 0.006 & 0.415 & 0.008 & 0.419 & 0.009 \\
\enddata
\tablenotetext{a}{Fluxes are normalized to 0.000701 photons cm$^{-2}$ s$^{-1}$.}
\end{deluxetable}

\startlongtable
\begin{deluxetable}{l|cc|cc|cc}
\tablecolumns{7}
\tablewidth{0pc}
\tablecaption{{\em 2XMM} Concordance Fluxes -- Hard Band\tablenotemark{a} \label{tab:2xmmhard} }
\tablehead{
\colhead{Target} & \multicolumn{2}{c}{pn} & \multicolumn{2}{c}{MOS1} & \multicolumn{2}{c}{MOS2} \\
\colhead{} & \colhead{$f_{ij}$} & \colhead{$\sigma_{ij}$} & \colhead{$f_{ij}$} & \colhead{$\sigma_{ij}$} & \colhead{$f_{ij}$} & \colhead{$\sigma_{ij}$} }
\startdata
            1127-145 & 0.173 & 0.007 & 0.186 & 0.008 & 0.179 & 0.008 \\ 
          1E0919+515 & 0.011 & 0.003 & 0.012 & 0.003 & 0.011 & 0.002 \\ 
             4C06.41 & 0.042 & 0.001 & 0.045 & 0.002 & 0.044 & 0.002 \\ 
       APM08279+5255 & 0.016 & 0.001 & 0.017 & 0.002 & 0.016 & 0.001 \\ 
              Cen X-4 & 0.018 & 0.003 & 0.018 & 0.003 & 0.019 & 0.003 \\ 
         CoD-33 7795 & 0.018 & 0.008 & 0.018 & 0.008 & 0.019 & 0.008 \\ 
         ESO323-G077 & 0.398 & 0.024 & 0.403 & 0.025 & 0.404 & 0.025 \\ 
           GRB080411 & 0.088 & 0.006 & 0.105 & 0.008 & 0.102 & 0.008 \\
\enddata
\tablenotetext{a}{Fluxes are normalized to 0.00223 photons cm$^{-2}$ s$^{-1}$.}
\end{deluxetable}

\startlongtable
\begin{deluxetable}{l|cc|cc|cc}
\tablecolumns{7}
\tablewidth{0pc}
\tablecaption{{\em XCAL} Concordance Fluxes -- Soft Band\tablenotemark{a} \label{tab:xcalsoft} }
\tablehead{
\colhead{Target} & \multicolumn{2}{c}{pn} & \multicolumn{2}{c}{MOS1} & \multicolumn{2}{c}{MOS2} \\
\colhead{} & \colhead{$f_{ij}$} & \colhead{$\sigma_{ij}$} & \colhead{$f_{ij}$} & \colhead{$\sigma_{ij}$} & \colhead{$f_{ij}$} & \colhead{$\sigma_{ij}$} }
\startdata
         1ES0414+009 & 0.0992 & 0.0100 & 0.1058 & 0.0107 & 0.1092 & 0.0113 \\ 
         1ES1101-232 & 0.2124 & 0.0058 & 0.2249 & 0.0063 & 0.2245 & 0.0064 \\ 
         1ES1553+113 & 0.1097 & 0.0033 & 0.1053 & 0.0032 & 0.1059 & 0.0032 \\ 
          1H0414+009 & 0.0459 & 0.0019 & 0.0477 & 0.0020 & 0.0491 & 0.0021 \\ 
          1H1219+301 & 0.2495 & 0.0140 & 0.2564 & 0.0145 & 0.2697 & 0.0154 \\ 
              3C 111 & 0.1780 & 0.0219 & 0.1883 & 0.0234 & 0.1893 & 0.0235 \\ 
              3C 111 & 0.1165 & 0.0081 & 0.1002 & 0.0070 & 0.1193 & 0.0083 \\ 
              3C 120 & 0.1999 & 0.0034 & 0.2112 & 0.0036 & 0.2079 & 0.0035 \\
\enddata
\tablenotetext{a}{Fluxes are normalized to 0.126 photons cm$^{-2}$ s$^{-1}$.}
\end{deluxetable}

\startlongtable
\begin{deluxetable}{l|cc|cc|cc}
\tablecolumns{7}
\tablewidth{0pc}
\tablecaption{{\em XCAL} Concordance Fluxes -- Medium Band\tablenotemark{a} \label{tab:xcalmedium} }
\tablehead{
\colhead{Target} & \multicolumn{2}{c}{pn} & \multicolumn{2}{c}{MOS1} & \multicolumn{2}{c}{MOS2} \\
\colhead{} & \colhead{$f_{ij}$} & \colhead{$\sigma_{ij}$} & \colhead{$f_{ij}$} & \colhead{$\sigma_{ij}$} & \colhead{$f_{ij}$} & \colhead{$\sigma_{ij}$} }
\startdata
         1ES0414+009 & 0.086 & 0.010 & 0.090 & 0.011 & 0.101 & 0.013 \\ 
         1ES1101-232 & 0.315 & 0.024 & 0.334 & 0.027 & 0.349 & 0.028 \\ 
         1ES1553+113 & 0.133 & 0.010 & 0.123 & 0.009 & 0.124 & 0.010 \\ 
          1H0414+009 & 0.044 & 0.006 & 0.045 & 0.006 & 0.046 & 0.007 \\ 
          1H1219+301 & 0.289 & 0.038 & 0.304 & 0.039 & 0.321 & 0.043 \\ 
              3C 111 & 0.361 & 0.043 & 0.397 & 0.047 & 0.393 & 0.047 \\ 
              3C 111 & 0.232 & 0.015 & 0.195 & 0.013 & 0.237 & 0.016 \\ 
              3C 120 & 0.222 & 0.034 & 0.232 & 0.037 & 0.238 & 0.038 \\
\enddata
\tablenotetext{a}{Fluxes are normalized to 0.0156 photons cm$^{-2}$ s$^{-1}$.}
\end{deluxetable}

\startlongtable
\begin{deluxetable}{l|cc|cc|cc}
\tablecolumns{7}
\tablewidth{0pc}
\tablecaption{{\em XCAL} Concordance Fluxes -- Hard Band\tablenotemark{a} \label{tab:xcalhard} }
\tablehead{
\colhead{Target} & \multicolumn{2}{c}{pn} & \multicolumn{2}{c}{MOS1} & \multicolumn{2}{c}{MOS2} \\
\colhead{} & \colhead{$f_{ij}$} & \colhead{$\sigma_{ij}$} & \colhead{$f_{ij}$} & \colhead{$\sigma_{ij}$} & \colhead{$f_{ij}$} & \colhead{$\sigma_{ij}$} }
\startdata
         1ES1101-232 & 0.1884 & 0.0062 & 0.1939 & 0.0070 & 0.2013 & 0.0077 \\ 
         1ES1553+113 & 0.0657 & 0.0025 & 0.0630 & 0.0022 & 0.0635 & 0.0022 \\ 
          1H0414+009 & 0.0152 & 0.0004 & 0.0159 & 0.0005 & 0.0155 & 0.0005 \\ 
          1H1219+301 & 0.1409 & 0.0145 & 0.1499 & 0.0162 & 0.1586 & 0.0176 \\ 
              3C 111 & 0.2704 & 0.0085 & 0.2958 & 0.0094 & 0.2921 & 0.0093 \\ 
              3C 111 & 0.2110 & 0.0036 & 0.1860 & 0.0033 & 0.2216 & 0.0039 \\ 
              3C 120 & 0.1652 & 0.0028 & 0.1701 & 0.0042 & 0.1703 & 0.0047 \\ 
              3C 120 & 0.1910 & 0.0007 & 0.1953 & 0.0010 & 0.1955 & 0.0011 \\
\enddata
\tablenotetext{a}{Fluxes are normalized to 0.0154 photons cm$^{-2}$ s$^{-1}$.}
\end{deluxetable}

\startlongtable
\begin{deluxetable}{lccllrrrr}
\tablecolumns{9}
\tablecaption{Capella grating observations with \axaf\ \label{tab:capella_obs}}
\tablehead{
\colhead{Epoch} & \colhead{ObsID} & \colhead{Exposure} & \colhead{Detector} & \colhead{Grating} & \multicolumn4c{Line Fluxes [10$^{-13}$~erg~s$^{-1}$~cm$^{-2}$]} \\
\colhead{\hfil} & \colhead{\hfil} & \colhead{[ks]} & \colhead{\hfil} & \colhead{Arm} & \colhead{\NeX~$\lambda$12} & \colhead{\FeXVII~$\lambda$15} & \colhead{\FeXVII~$\lambda$17} & \colhead{\OVIII~$\lambda$19}
}
\startdata
Aug 1999 & 1099 & 14.57 & ACIS-S & HEG+1 & 21.1$\pm$2.0 & 57.8$\pm$4.7 &  ...  &  ...  \\ 
\multicolumn{4}{c}{\hfil} & HEG-1 & 21.8$\pm$1.7 & 63.3$\pm$4.8 & 64.3$\pm$6.4 & ... \\ 
\multicolumn{4}{c}{\hfil} & MEG+1 & 21.0$\pm$1.0 & 47.5$\pm$2.1 & 68.9$\pm$3.3 & 29.3$\pm$2.8 \\
\multicolumn{4}{c}{\hfil} & MEG-1 & 20.0$\pm$1.0 & 48.7$\pm$1.5 & 69.4$\pm$2.1 & 34.4$\pm$1.7 \\

Aug 1999 & 1235 & 14.57 & ACIS-S & HEG+1 & 22.4$\pm$2.1 & 59.1$\pm$4.7 &  ...  &  ...  \\ 
\multicolumn{4}{c}{\hfil} & HEG-1 & 20.5$\pm$1.6 & 47.2$\pm$4.2 & 66.3$\pm$6.4 & 31.3$\pm$14.4 \\
\multicolumn{4}{c}{\hfil} & MEG+1 & 20.5$\pm$1.0 & 50.7$\pm$2.2 & 68.0$\pm$3.3 & 30.7$\pm$2.9 \\
\multicolumn{4}{c}{\hfil} & MEG-1 & 20.6$\pm$1.0 & 51.7$\pm$1.6 & 69.9$\pm$2.1 & 33.3$\pm$1.7 \\
\enddata
\end{deluxetable}

\bibliographystyle{aasjournal}                       
\bibliography{apj-jour,concordance}

\begin{thebibliography}{}
\expandafter\ifx\csname natexlab\endcsname\relax\def\natexlab#1{#1}\fi
\providecommand{\url}[1]{\href{#1}{#1}}

\bibitem[{{Beuermann} {et~al.}(2006){Beuermann}, {Burwitz}, \&
  {Rauch}}]{2006A&A...458..541B}
{Beuermann}, K., {Burwitz}, V., \& {Rauch}, T. 2006, \aap, 458, 541

\bibitem[{Chen {et~al.}(2019)Chen, Meng, Wang, van Dyk, Marshall, \&
  Kashyap}]{concordancejasa}
Chen, Y., Meng, X.-L., Wang, X., {et~al.} 2019, Journal of the American
  Statistical Association, 114, 1018.
\newblock \url{https://doi.org/10.1080/01621459.2018.1528978}

\bibitem[{{Drake} {et~al.}(2006){Drake}, {Ratzlaff}, {Kashyap}, {Edgar},
  {Izem}, {Jerius}, {Siemiginowska}, \& {Vikhlinin}}]{2006SPIE.6270E..1ID}
{Drake}, J.~J., {Ratzlaff}, P., {Kashyap}, V., {et~al.} 2006, in \procspie,
  Vol. 6270, Society of Photo-Optical Instrumentation Engineers (SPIE)
  Conference Series, 62701I

\bibitem[{Efron \& Morris(1972)}]{efron1972empirical}
Efron, B., \& Morris, C. 1972, Biometrika, 59, 335

\bibitem[{Efron \& Morris(1973)}]{efron1973stein}
---. 1973, Journal of the American Statistical Association, 68, 117

\bibitem[{{Fruscione} {et~al.}(2006){Fruscione}, {McDowell}, {Allen},
  {Brickhouse}, {Burke}, {Davis}, {Durham}, {Elvis}, {Galle}, {Harris},
  {Huenemoerder}, {Houck}, {Ishibashi}, {Karovska}, {Nicastro}, {Noble},
  {Nowak}, {Primini}, {Siemiginowska}, {Smith}, \&
  {Wise}}]{2006SPIE.6270E..1VF}
{Fruscione}, A., {McDowell}, J.~C., {Allen}, G.~E., {et~al.} 2006, in
  \procspie, Vol. 6270, Society of Photo-Optical Instrumentation Engineers
  (SPIE) Conference Series, 62701V

\bibitem[{{Gabriel} {et~al.}(2004){Gabriel}, {Denby}, {Fyfe}, {Hoar}, {Ibarra},
  {Ojero}, {Osborne}, {Saxton}, {Lammers}, \& {Vacanti}}]{2004ASPC..314..759G}
{Gabriel}, C., {Denby}, M., {Fyfe}, D.~J., {et~al.} 2004, in Astronomical
  Society of the Pacific Conference Series, Vol. 314, Astronomical Data
  Analysis Software and Systems (ADASS) XIII, ed. F.~{Ochsenbein}, M.~G.
  {Allen}, \& D.~{Egret}, 759

\bibitem[{{Guainazzi} {et~al.}(2015){Guainazzi}, {David}, {Grant}, {Miller},
  {Natalucci}, {Nevalainen}, {Petre}, \& {Audard}}]{2015JATIS...1d7001G}
{Guainazzi}, M., {David}, L., {Grant}, C.~E., {et~al.} 2015, Journal of
  Astronomical Telescopes, Instruments, and Systems, 1, 047001

\bibitem[{{Ishida} {et~al.}(2011){Ishida}, {Tsujimoto}, {Kohmura},
  {Stuhlinger}, {Smith}, {Marshall}, {Guainazzi}, {Kawai}, \&
  {Ogawa}}]{2011PASJ...63S.657I}
{Ishida}, M., {Tsujimoto}, M., {Kohmura}, T., {et~al.} 2011, \pasj, 63, S657

\bibitem[{{Jansen} {et~al.}(2001){Jansen}, {Lumb}, {Altieri}, {Clavel}, {Ehle},
  {Erd}, {Gabriel}, {Guainazzi}, {Gondoin}, {Much}, {Munoz}, {Santos},
  {Schartel}, {Texier}, \& {Vacanti}}]{2001A&A...365L...1J}
{Jansen}, F., {Lumb}, D., {Altieri}, B., {et~al.} 2001, \aap, 365, L1

\bibitem[{{Jethwa} {et~al.}(2015){Jethwa}, {Saxton}, {Guainazzi},
  {Rodriguez-Pascual}, \& {Stuhlinger}}]{2015A&A...581A.104J}
{Jethwa}, P., {Saxton}, R., {Guainazzi}, M., {Rodriguez-Pascual}, P., \&
  {Stuhlinger}, M. 2015, \aap, 581, A104

\bibitem[{{Kashyap} \& {Drake}(2000)}]{2000BASI...28..475K}
{Kashyap}, V., \& {Drake}, J.~J. 2000, Bulletin of the Astronomical Society of
  India, 28, 475

\bibitem[{{Kashyap} {et~al.}(2008){Kashyap}, {Lee}, {Siemiginowska},
  {McDowell}, {Rots}, {Drake}, {Ratzlaff}, {Zezas}, {Izem}, {Connors}, {van
  Dyk}, \& {Park}}]{2008SPIE.7016E..0PK}
{Kashyap}, V.~L., {Lee}, H., {Siemiginowska}, A., {et~al.} 2008, in Society of
  Photo-Optical Instrumentation Engineers (SPIE) Conference Series, Vol. 7016,
  Observatory Operations: Strategies, Processes, and Systems II, ed. R.~J.
  {Brissenden} \& D.~R. {Silva}, 70160P

\bibitem[{{Kettula} {et~al.}(2013){Kettula}, {Nevalainen}, \&
  {Miller}}]{2013A&A...552A..47K}
{Kettula}, K., {Nevalainen}, J., \& {Miller}, E.~D. 2013, \aap, 552, A47

\bibitem[{{Kirsch} {et~al.}(2005){Kirsch}, {Briel}, {Burrows}, {Campana},
  {Cusumano}, {Ebisawa}, {Freyberg}, {Guainazzi}, {Haberl}, {Jahoda},
  {Kaastra}, {Kretschmar}, {Larsson}, {Lubi{\'n}ski}, {Mori}, {Plucinsky},
  {Pollock}, {Rothschild}, {Sembay}, {Wilms}, \&
  {Yamamoto}}]{2005SPIE.5898...22K}
{Kirsch}, M.~G., {Briel}, U.~G., {Burrows}, D., {et~al.} 2005, in Society of
  Photo-Optical Instrumentation Engineers (SPIE) Conference Series, Vol. 5898,
  \procspie, ed. O.~H.~W. {Siegmund}, 22--33

\bibitem[{{Lee} {et~al.}(2011){Lee}, {Kashyap}, {van Dyk}, {Connors}, {Drake},
  {Izem}, {Meng}, {Min}, {Park}, {Ratzlaff}, {Siemiginowska}, \&
  {Zezas}}]{2011ApJ...731..126L}
{Lee}, H., {Kashyap}, V.~L., {van Dyk}, D.~A., {et~al.} 2011, \apj, 731, 126

\bibitem[{{Madsen} {et~al.}(2017{\natexlab{a}}){Madsen}, {Beardmore},
  {Forster}, {Guainazzi}, {Marshall}, {Miller}, {Page}, \&
  {Stuhlinger}}]{2017AJ....153....2M}
{Madsen}, K.~K., {Beardmore}, A.~P., {Forster}, K., {et~al.}
  2017{\natexlab{a}}, \aj, 153, 2

\bibitem[{{Madsen} {et~al.}(2017{\natexlab{b}}){Madsen}, {Forster},
  {Grefenstette}, {Harrison}, \& {Stern}}]{2017ApJ...841...56M}
{Madsen}, K.~K., {Forster}, K., {Grefenstette}, B.~W., {Harrison}, F.~A., \&
  {Stern}, D. 2017{\natexlab{b}}, \apj, 841, 56

\bibitem[{{Madsen} {et~al.}(2019){Madsen}, {Natalucci}, {Belanger}, {Grant},
  {Guainazzi}, {Kashyap}, {Marshall}, {Miller}, {Nevalainen}, {Plucinsky}, \&
  {Terada}}]{iachec2019}
{Madsen}, K.~K., {Natalucci}, L., {Belanger}, G., {et~al.} 2019, arXiv
  e-prints, arXiv:1901.00934

\bibitem[{{Madsen} {et~al.}(2020){Madsen}, {Terada}, {Burwitz}, {Belanger},
  {Grant}, {Guainazzi}, {Kashyap}, {Marshall}, {Miller}, {Natalucci}, \&
  {Plucinsky}}]{iachec2020}
{Madsen}, K.~K., {Terada}, Y., {Burwitz}, V., {et~al.} 2020, arXiv e-prints,
  arXiv:2001.11117

\bibitem[{{Nevalainen} {et~al.}(2010){Nevalainen}, {David}, \&
  {Guainazzi}}]{2010A&A...523A..22N}
{Nevalainen}, J., {David}, L., \& {Guainazzi}, M. 2010, \aap, 523, A22

\bibitem[{{Plucinsky} {et~al.}(2017){Plucinsky}, {Beardmore}, {Foster},
  {Haberl}, {Miller}, {Pollock}, \& {Sembay}}]{2017A&A...597A..35P}
{Plucinsky}, P.~P., {Beardmore}, A.~P., {Foster}, A., {et~al.} 2017, \aap, 597,
  A35

\bibitem[{{Read} {et~al.}(2014){Read}, {Guainazzi}, \&
  {Sembay}}]{2014A&A...564A..75R}
{Read}, A.~M., {Guainazzi}, M., \& {Sembay}, S. 2014, \aap, 564, A75

\bibitem[{{Schellenberger} {et~al.}(2015){Schellenberger}, {Reiprich},
  {Lovisari}, {Nevalainen}, \& {David}}]{2015A&A...575A..30S}
{Schellenberger}, G., {Reiprich}, T.~H., {Lovisari}, L., {Nevalainen}, J., \&
  {David}, L. 2015, \aap, 575, A30

\bibitem[{{Str{\"u}der} {et~al.}(2001){Str{\"u}der}, {Briel}, {Dennerl},
  {Hartmann}, {Kendziorra}, {Meidinger}, {Pfeffermann}, {Reppin}, {Aschenbach},
  {Bornemann}, {Br{\"a}uninger}, {Burkert}, {Elender}, {Freyberg}, {Haberl},
  {Hartner}, {Heuschmann}, {Hippmann}, {Kastelic}, {Kemmer}, {Kettenring},
  {Kink}, {Krause}, {M{\"u}ller}, {Oppitz}, {Pietsch}, {Popp}, {Predehl},
  {Read}, {Stephan}, {St{\"o}tter}, {Tr{\"u}mper}, {Holl}, {Kemmer}, {Soltau},
  {St{\"o}tter}, {Weber}, {Weichert}, {von Zanthier}, {Carathanassis}, {Lutz},
  {Richter}, {Solc}, {B{\"o}ttcher}, {Kuster}, {Staubert}, {Abbey}, {Holland},
  {Turner}, {Balasini}, {Bignami}, {La Palombara}, {Villa}, {Buttler},
  {Gianini}, {Lain{\'e}}, {Lumb}, \& {Dhez}}]{2001A&A...365L..18S}
{Str{\"u}der}, L., {Briel}, U., {Dennerl}, K., {et~al.} 2001, \aap, 365, L18

\bibitem[{{Toor} \& {Seward}(1974)}]{1974AJ.....79..995T}
{Toor}, A., \& {Seward}, F.~D. 1974, \aj, 79, 995

\bibitem[{{Tsujimoto} {et~al.}(2011){Tsujimoto}, {Guainazzi}, {Plucinsky},
  {Beardmore}, {Ishida}, {Natalucci}, {Posson-Brown}, {Read}, {Saxton}, \&
  {Shaposhnikov}}]{2011A&A...525A..25T}
{Tsujimoto}, M., {Guainazzi}, M., {Plucinsky}, P.~P., {et~al.} 2011, \aap, 525,
  A25

\bibitem[{{Turner} {et~al.}(2001){Turner}, {Abbey}, {Arnaud}, {Balasini},
  {Barbera}, {Belsole}, {Bennie}, {Bernard}, {Bignami}, {Boer}, {Briel},
  {Butler}, {Cara}, {Chabaud}, {Cole}, {Collura}, {Conte}, {Cros}, {Denby},
  {Dhez}, {Di Coco}, {Dowson}, {Ferrando}, {Ghizzardi}, {Gianotti}, {Goodall},
  {Gretton}, {Griffiths}, {Hainaut}, {Hochedez}, {Holland}, {Jourdain},
  {Kendziorra}, {Lagostina}, {Laine}, {La Palombara}, {Lortholary}, {Lumb},
  {Marty}, {Molendi}, {Pigot}, {Poindron}, {Pounds}, {Reeves}, {Reppin},
  {Rothenflug}, {Salvetat}, {Sauvageot}, {Schmitt}, {Sembay}, {Short},
  {Spragg}, {Stephen}, {Str{\"u}der}, {Tiengo}, {Trifoglio}, {Tr{\"u}mper},
  {Vercellone}, {Vigroux}, {Villa}, {Ward}, {Whitehead}, \&
  {Zonca}}]{2001A&A...365L..27T}
{Turner}, M.~J.~L., {Abbey}, A., {Arnaud}, M., {et~al.} 2001, \aap, 365, L27

\bibitem[{{Watson} {et~al.}(2009){Watson}, {Schr{\"o}der}, {Fyfe}, {Page},
  {Lamer}, {Mateos}, {Pye}, {Sakano}, {Rosen}, {Ballet}, {Barcons}, {Barret},
  {Boller}, {Brunner}, {Brusa}, {Caccianiga}, {Carrera}, {Ceballos}, {Della
  Ceca}, {Denby}, {Denkinson}, {Dupuy}, {Farrell}, {Fraschetti}, {Freyberg},
  {Guillout}, {Hambaryan}, {Maccacaro}, {Mathiesen}, {McMahon}, {Michel},
  {Motch}, {Osborne}, {Page}, {Pakull}, {Pietsch}, {Saxton}, {Schwope},
  {Severgnini}, {Simpson}, {Sironi}, {Stewart}, {Stewart}, {Stobbart}, {Tedds},
  {Warwick}, {Webb}, {West}, {Worrall}, \& {Yuan}}]{2009A&A...493..339W}
{Watson}, M.~G., {Schr{\"o}der}, A.~C., {Fyfe}, D., {et~al.} 2009, \aap, 493,
  339

\bibitem[{{Xu} {et~al.}(2014){Xu}, {van Dyk}, {Kashyap}, {Siemiginowska},
  {Connors}, {Drake}, {Meng}, {Ratzlaff}, \& {Yu}}]{2014ApJ...794...97X}
{Xu}, J., {van Dyk}, D.~A., {Kashyap}, V.~L., {et~al.} 2014, \apj, 794, 97

\end{thebibliography}

\end{document}